\RequirePackage[l2tabu,orthodox]{nag}

\documentclass[headsepline,footsepline,footinclude=false,oneside,fontsize=11pt,paper=a4,listof=totoc,bibliography=totoc]{scrbook} 

\PassOptionsToPackage{table,svgnames,dvipsnames}{xcolor}

\usepackage[utf8]{inputenc}
\usepackage[T1]{fontenc}
\usepackage[sc]{mathpazo}
\usepackage[ngerman,american]{babel}
\usepackage[autostyle]{csquotes}
\usepackage[%
  backend=biber,
  url=false,
  style=numeric,
  maxnames=4,
  minnames=3,
  maxbibnames=99,
  giveninits,
  uniquename=init,
  sorting=none]{biblatex} 
\usepackage{graphicx}
\usepackage{scrhack} 
\usepackage{listings}
\usepackage{lstautogobble}
\usepackage{tikz}
\usepackage{pgfplots}
\usepackage{pgfplotstable}
\usepackage{booktabs}
\usepackage[final]{microtype}
\usepackage{caption}
\usepackage[hidelinks]{hyperref} 

\usepackage{physics}
\usepackage{siunitx}
\usepackage[compat=1.1.0]{tikz-feynhand}
\usepackage{subcaption}
\usepackage{float}
\usepackage{amssymb}

\usepackage{amsmath}
\usepackage{todonotes}

\setkomafont{disposition}{\normalfont\bfseries} 
\BeforeTOCHead[toc]{{\cleardoublepage\pdfbookmark[0]{\contentsname}{toc}}}

\definecolor{TUMBlue}{HTML}{0065BD}
\definecolor{TUMSecondaryBlue}{HTML}{005293}
\definecolor{TUMSecondaryBlue2}{HTML}{003359}
\definecolor{TUMBlack}{HTML}{000000}
\definecolor{TUMWhite}{HTML}{FFFFFF}
\definecolor{TUMDarkGray}{HTML}{333333}
\definecolor{TUMGray}{HTML}{808080}
\definecolor{TUMLightGray}{HTML}{CCCCC6}
\definecolor{TUMAccentGray}{HTML}{DAD7CB}
\definecolor{TUMAccentOrange}{HTML}{E37222}
\definecolor{TUMAccentGreen}{HTML}{A2AD00}
\definecolor{TUMAccentLightBlue}{HTML}{98C6EA}
\definecolor{TUMAccentBlue}{HTML}{64A0C8}

\pgfplotsset{compat=newest}
\pgfplotsset{
  cycle list={TUMBlue\\TUMAccentOrange\\TUMAccentGreen\\TUMSecondaryBlue2\\TUMDarkGray\\},
}

\newcommand*{\getTitle}{Density-Dependent Neutron-Neutron Interaction from Subleading Chiral Three-Neutron Forces}

\usepackage[titletoc]{appendix}

\begin{document}


\pagenumbering{alph}
{

\oddsidemargin=\evensidemargin\relax
\textwidth=\dimexpr\paperwidth-2\evensidemargin-2in\relax
\hsize=\textwidth\relax

\begin{center}
  {\huge{\bfseries \getTitle{}\,\footnote{\,Bachelor's thesis in physics, Technische Universit\"at M\"unchen, September 2020}
  }  }  

\bigskip
\bigskip

 Lukas Treuer \\
\bigskip
{Physik-Department T39, Technische Universit\"at M\"unchen,
   85748 Garching b. M\"unchen, Germany
   
\bigskip

{\itshape E-Mail: lukas.treuer@tum.de}}
\end{center}

\bigskip
\bigskip

\begin{center}
{\Large{\bfseries \abstractname}}
\end{center}
\medskip

Three-nucleon forces are an essential ingredient for an accurate description of nuclear few- and many-body systems. However, implementing them directly in many-body calculations is technically very challenging. Thus, there is a need for an efficient approximation method. By closing one nucleon line to a loop, it is possible to derive effective in-medium nucleon-nucleon interactions that represent the underlying three-nucleon forces, as constructed in Chiral Effective Field Theory. Since three-neutron forces are equally as important for the computation of the equation of state for pure neutron matter, this work applies the aforementioned approach to the subleading chiral three-neutron forces, in particular the short-range terms and relativistic corrections. It is shown in this work that, while many contributions to the in-medium neutron-neutron interaction are - apart from a constant factor - identical to the terms in isospin-symmetric matter, some differ drastically. Moreover, previously vanishing terms yield now non-zero contributions. As a result of this work, density-dependent in-medium neutron-neutron potentials are now available for the implementation in nuclear many-body calculations, either in closed analytical form, or requiring at most one numerical integration.

\begin{center}
    \IfFileExists{blank.png}{%
    \vfill{}
    \includegraphics[height=0mm]{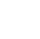}
  }{}
\end{center} 

}
\thispagestyle{empty}

\frontmatter{}

\microtypesetup{protrusion=false}
\tableofcontents{}
\thispagestyle{empty}
\microtypesetup{protrusion=true}

\mainmatter{}


\chapter{Introduction}\label{chapter:introduction}

Three-nucleon (3N) forces allow for a deeper understanding of the strong interactions and they provide more precise predictions for nuclear many-body systems, thus playing an important role not only in nuclear physics, but in astrophysics as well. Particularly in nuclear few-body systems, 3N-forces represent an essential ingredient for the accurate description of experimental data and basic nuclear phenomena, such as binding energy per particle or the saturation density of nuclear matter \cite{LO}. \\
The 3N-interactions are constructed using Chiral Effective Field Theory (ChEFT), which governs the low-energy regime of the quantum field theory of the fundamental strong interaction, namely Quantum Chromo Dynamics. This is achieved by making use of spontaneous and explicit breaking of chiral symmetry, induced by a non-zero scalar quark condensate and non-vanishing current quark masses, respectively. Thus, the relevant degrees of freedom are the nucleons on the one hand side, and the Goldstone bosons of the spontaneously broken chiral symmetry on the other hand side. The latter comprise the light pseudoscalar meson octet, or the tree pions in the case of two quark-flavours \cite{LO,forces}. \\
In the low-momentum expansion of chiral EFT, 3N-forces are not present up to next-to-next-to-leading order (N$^2$LO). Up to that order, the NN-interaction consists of effective zero-range contact terms and longer-range components through one-pion- ($1\pi$) and two-pion- ($2\pi$) exchanges \cite{LO}. \\\

When implementing 3N-forces in nuclear many-body systems, however, it is computationally very difficult to calculate their contributions directly. Therefore, a simpler method was developed in ref. \cite{LO}, where the inclusion of 3N-forces is done via a density-dependent nucleon-nucleon (NN) interaction. Since this involves the construction of an effective in-medium potential $V_{med}$\,, it is only a good approximation when dealing with many-body systems that can be viewed in the thermodynamic limit. The systems are assumed to be at zero temperature, thus restricting the occupied states of nucleons to below the Fermi surface. \\
The approach of constructing $V_{med}$ from chiral 3N-forces was not only implemented for the leading chiral 3N-forces \cite{LO}, but for the subleading contributions, as well. The subleading 3N-forces were divided into short-range terms and relativistic corrections \cite{sr}, intermediate-range terms \cite{IR}, as well as long-range terms \cite{LR}. Specifically, in the mentioned articles, the method of consructing $V_{med}$ was executed for isospin-symmetric spin-saturated nuclear matter. The results are thus applicable to nuclear systems with the same number of protons and neutrons, and fully paired spins. \\\

However, up to that point, the contributions of subleading chiral 3n-forces to an effective in-medium interaction in pure neutron matter had not been calculated. Hence, the aim of this work is to begin the application of the previously mentioned approximation method to spin-saturated neutron matter. Specifically, this work presents the results for density-dependent neutron-neutron (nn) interactions arising from short range terms and relativistic corrections of the subleading chiral three-neutron (3n) forces, originally derived in ref. \cite{forces} for three nucleons. \\
Adapting them accordingly to represent only three-neutron interactions, the subleading chiral 3N-forces used as starting point are taken from ref. \cite{sr} for the sake of consistency and comparability, as some corrections and specifications have been made with respect to ref. \cite{forces}. \\
In perspective, the results presented in this work will be useful for implementation in nuclear many-body calculations, in order to deepen our understanding of systems such as neutron stars, as well as to gain more insight into some physical parameters of nuclear matter, e.g. the isospin-asymmetry energy and its slope as a function of nuclear density. These can be obtained from the equation of state of pure neutron matter and symmetric nuclear matter \cite{EOS}. \\\

Beginning the main section of this work, the topologies involved in the interaction, as well as the method of calculation are detailed in chapter \ref{chapter:topologies}. The results for in-medium nn-potentials are presented together with the 3n-forces they stem from in chapter \ref{chapter:results}. Following the accompanying discussions, a summarizing conclusion and an outlook to future studies are given in chapter \ref{chapter:conclusion}. Finally, all the relevant loop-functions are defined and specified in the appendix, either in closed analytical form, or through one-parameter integrals.

\chapter{Topologies of One-Loop Diagrams and the In-Medium Neutron-Neutron Interaction}\label{chapter:topologies}

In order to discuss the effective in-medium potentials presented in chapter \ref{chapter:results}, it is helpful to first introduce the general form of the 3n-interaction and to understand the involved topologies, as well as the method of calculation. To that end, thorough explanations are given in this chapter. That includes the derivation of the basic identities characterizing the situation in pure neutron matter, related to isospin operators. \\
For the sake of comprehensibility, this chapter is divided into two sections:\\
Section (\ref{sec:topo}) deals with the fundamental interaction scheme of the considered 3n-forces and the subsequently appearing in-medium topologies, while the technicalities of deriving the effective nn-potentials are explained in section (\ref{sec:method}).

\section{Generic Three-Neutron Interaction and One-Loop Topologies}\label{sec:topo}

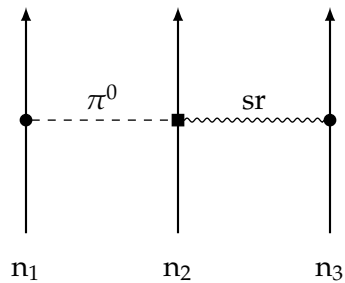
\begin{figure}[htbp]
	\centering
	\begin{tikzpicture}
		\begin{feynhand}
			\vertex[dot] (b1) at (0,1.5) {};
			\draw[->,>=latex, thick] (0,0) to (0,3);
			\node at (0,-0.5) {n$_1$};
			\vertex[squaredot] (b2) at (2,1.5) {};
			\draw[->,>=latex, thick] (2,0) to (2,3);
			\node at (2,-0.5) {n$_2$};
			\vertex[dot] (b3) at (4,1.5) {};
			\draw[->,>=latex, thick] (4,0) to (4,3);
			\node at (4,-0.5) {n$_3$};
			\propag[sca, thick] (b1) to [edge label = $\pi^0$](b2);
			\propag[bos, thick] (b2) to [edge label = \vphantom{$\pi^0$}sr] (b3);
		\end{feynhand}
	\end{tikzpicture}
	\caption{Generic 3n-interaction: $\pi^0$-exchange (dashed line) between the first and second neutron, short-range interaction (wiggly line) between neutron 2 and neutron 3}
	\label{fig:generic}
\end{figure}

Beginning with the generic form of the 3n-interaction shown in figure (\ref{fig:generic}), it is important to familiarize oneself with the relevant interaction mechanism. The three neutrons interact via the exchange of one pion (or two pions) and an effective zero-range contact coupling containing a momentum-independent propagator. Also, there exist cases where the symbolic diagram in fig. (\ref{fig:generic}) is interpreted as two-pion-exchange. In particular, this applies to the 3n-interactions presented in subsection (\ref{ssec:r2p}). \\
However, contrary to 3N-forces, only the neutral pion $\pi^0$ - within the pion-triplet - is involved in 3n-forces. In other words, the strong nuclear interaction in isospin-symmetric matter can be executed via the exchange of neutral and charged pions, whereas only the neutral pion can mediate the residual strong interaction between neutrons, due to charge conservation. \\\

\begin{figure}[H]
	\centering
	\begin{subfigure}{.4\textwidth}
		\centering
		\begin{tikzpicture}
		\begin{feynhand}
			\vertex[squaredot] (b1) at (0,1.5) {};
			\draw[->,>=latex, thick] (0,0) to (0,3);
			\node at (0,-0.5) {n$_1$};
			\vertex[color=black!0,dot] (b2) at (1.3,1.5) {};
			\draw[color=black!0,->,>=latex, thick] (1.3,0) to (1.3,3);
			\node[color=black!0] at (1.3,-0.5) {n$_2$};
			\vertex[color=black!0,dot] (b3) at (2.6,1.5) {};
			\draw[->,>=latex, thick] (2.6,0) to (2.6,3);
			\node at (2.6,-0.5) {n$_2$};
			\draw[-, thick] (2.45,1.5) to (2.75,1.7);
			\draw[-, thick] (2.45,1.3) to (2.75,1.5);
			\vertex[dot] (p) at (2.6,2.3){};
			\vertex[dot] (sr) at (2.6,0.7){};
			\propag[sca, thick] (b1) to [edge label = $\pi^0$](p);
			\propag[bos, thick] (b1) to [edge label' = \vphantom{$\pi^0$}sr] (sr);
		\end{feynhand}
		\end{tikzpicture}
	\end{subfigure}
	\begin{subfigure}{.4\textwidth}
		\centering
		\begin{tikzpicture}
		\begin{feynhand}
			\vertex[squaredot] (b1) at (0,1.5) {};
			\draw[->,>=latex, thick] (0,0) to (0,3);
			\node at (0,-0.5) {n$_1$};
			\vertex[color=black!0,dot] (b2) at (1.3,1.5) {};
			\draw[color=black!0,->,>=latex, thick] (1.3,0) to (1.3,3);
			\node[color=black!0] at (1.3,-0.5) {n$_2$};
			\vertex[color=black!0,dot] (b3) at (2.6,1.5) {};
			\draw[->,>=latex, thick] (2.6,0) to (2.6,3);
			\node at (2.6,-0.5) {n$_2$};
			\draw[-, thick] (2.45,1.5) to (2.75,1.7);
			\draw[-, thick] (2.45,1.3) to (2.75,1.5);
			\vertex[dot] (sr) at (2.6,2.3){};
			\vertex[dot] (p) at (2.6,0.7){};
			\propag[sca, thick] (b1) to [edge label' = $\pi^0$](p);
			\propag[bos, thick] (b1) to [edge label = \vphantom{$\pi^0$}sr] (sr);
		\end{feynhand}
		\end{tikzpicture}
	\end{subfigure}
	\caption{Feynman diagrams from double exchanges yielding $V_{med}^{(3)}$\,. Graphs arising from the exchange $n_1 \leftrightarrow n_2$ are not shown and have to be added.}
	\label{fig:de}
\end{figure}
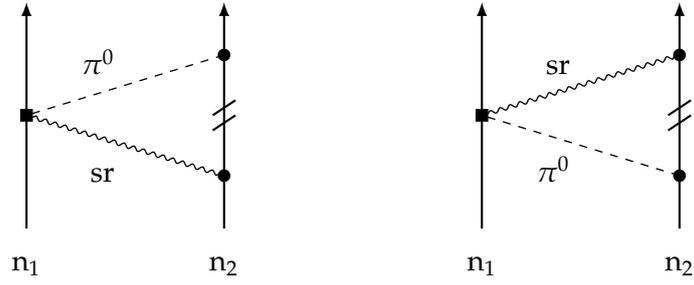

After considering the generic 3n-interaction, the focus now lies on the appearing topologies when calculating the in-medium potential. Their diagrammatic structure is shown in figs. (\ref{fig:sc}) - (\ref{fig:de}), where mirror images - in other words contributions arising from the exchange of the external neutron lines $n_1 \leftrightarrow n_2$ - have to be added. In some cases this simply yields a factor of 2. Thus, the relevant topologies are introduced here, whereas their computational intricacies are detailed in the following section. \\
Beginning with the term "topology" itself, whose meaning somewhat differs depending on the field and topic. It will henceforth denote the equivalent diagrammatic structure or meaning of Feynman diagrams, and as such the underlying terms in the transition amplitude, constructed from the Feynman rules of the appropriate quantum field theory - in this case ChEFT. \\
Fundamentally, one obtains the in-medium nn-potential by closing one of the three neutron lines to an in-medium loop, leading to the four distinct topologies. When doing this, it is critical to follow the appropriate Feynman rules, that is, integrating over the respective neutron four-momentum, and in the case of fermionic loops, taking the trace over the neutron's spin states. Additionally, there appears a minus-sign for closed fermion lines. \\\

First of all, it is possible to close one neutron line to itself, which can be seen in fig. (\ref{fig:sc}). Such contributions from self-closings are henceforth denoted as $V_{med}^{(0)}$\,. \\
Secondly, one obtains two different kinds of vertex corrections, through a short-range interaction on the one hand, and through pion exchange on the other hand, giving rise to the pieces $V_{med}^{(1)}$ and $V_{med}^{(2)}$\,, respectively. These topologies are visible in figs. (\ref{fig:srvc}) and (\ref{fig:pvc}). \\
Lastly, the fourth topology describes double exchanges, whose structure is shown in fig. (\ref{fig:de}), yielding the contribution $V_{med}^{(3)}$\,. \\\

\begin{figure}[H]
	\centering
	\begin{subfigure}{.3\textwidth}
		\centering
		\begin{tikzpicture}
			\begin{feynhand}
				\vertex[dot] (b1) at (0.6,1.5) {};
				\vertex[squaredot] (b2) at (1.5,1.5) {};
				\draw[->, >=latex, thick] (1.5,0) to (1.5,3);
				\node at (1.5,-0.5) {n$_1$};
				\propag[fer, ultra thick] (b1) to [half right, looseness=1.65] (-0.6,1.5);
				\propag[fer, ultra thick] (-0.6,1.5) to [half right, looseness=1.65] (b1);
				\draw[-, thick] (-0.75,1.5) to (-0.45,1.7);
				\draw[-, thick] (-0.75,1.3) to (-0.45,1.5);
				\vertex[dot] (b3) at (3,1.5) {};
				\draw[->, >=latex, thick] (3,0) to (3,3);
				\node at (3,-0.5) {n$_2$};
				\propag[sca, thick] (b1) to [edge label' = $\pi^0$](b2);
				\propag[bos, thick] (b2) to [edge label' = \vphantom{$\pi^0$}sr ] (b3);
			\end{feynhand}
		\end{tikzpicture}
	\end{subfigure}
	\begin{subfigure}{.3\textwidth}
		\centering
		\begin{tikzpicture}
			\begin{feynhand}
				\vertex[dot] (b1) at (0,1.5) {};
				\draw[->,>=latex, thick] (0,0) to (0,3);
				\node at (0,-0.5) {n$_1$};
				\vertex[squaredot] (b2) at (1.5,1.5) {};
				\propag[fer, ultra thick] (b2) to [half right, looseness=1.6] (1.5,2.7);
				\propag[fer, ultra thick] (1.5,2.7) to [half right, looseness=1.6] (b2);
				\draw[-, thick] (1.3,2.55) to (1.5,2.85);
				\draw[-, thick] (1.5,2.55) to (1.7,2.85);
				\vertex[dot] (b3) at (3,1.5) {};
				\draw[->,>=latex, thick] (3,0) to (3,3);
				\node at (3,-0.5) {n$_2$};
				\propag[sca, thick] (b1) to [edge label' = $\pi^0$](b2);
				\propag[bos, thick] (b2) to [edge label' = \vphantom{$\pi^0$}sr ] (b3);
			\end{feynhand}
		\end{tikzpicture}
	\end{subfigure}
	\begin{subfigure}{.3\textwidth}
		\centering
		\begin{tikzpicture}
			\begin{feynhand}
				\vertex[dot] (b1) at (0,1.5) {};
				\draw[->,>=latex, thick] (0,0) to (0,3);
				\node at (0,-0.5) {n$_1$};
				\vertex[squaredot] (b2) at (1.5,1.5) {};
				\draw[->,>=latex, thick] (1.5,0) to (1.5,3);
				\node at (1.5,-0.5) {n$_2$};
				\vertex[dot] (b3) at (2.4,1.5) {};
				\propag[fer, ultra thick] (b3) to [half right, looseness=1.65] (3.6,1.5);
				\propag[fer, ultra thick] (3.6,1.5) to [half right, looseness=1.65] (b3);
				\draw[-, thick] (3.45,1.5) to (3.75,1.7);
				\draw[-, thick] (3.45,1.3) to (3.75,1.5);
				\propag[sca, thick] (b1) to [edge label' = $\pi^0$](b2);
				\propag[bos, thick] (b2) to [edge label' = sr \vphantom{$\pi^0$}] (b3);
			\end{feynhand}
		\end{tikzpicture}
	\end{subfigure}
	\caption{Feynman diagrams from self-closings yielding $V_{med}^{(0)}$\,. Graphs arising from the exchange $n_1 \leftrightarrow n_2$ are not shown and have to be added.}
	\label{fig:sc}
\end{figure}
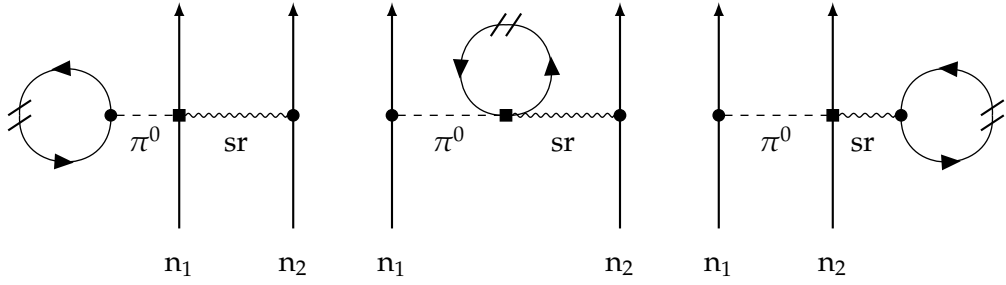

\begin{figure}[H]
	\centering
	\begin{subfigure}{.4\textwidth}
		\centering
		\begin{tikzpicture}
		\begin{feynhand}
			\vertex[dot] (b1) at (0,1.5) {};
			\draw[->,>=latex, thick] (0,0) to (0,3);
			\node at (0,-0.5) {n$_1$};
			\vertex[squaredot] (b2) at (1.75,1.5) {};
			\draw[->,>=latex, thick] (1.75,0) to (1.75,3);
			\draw[-, thick] (1.6,2) to (1.9,2.2);
			\draw[-, thick] (1.6,1.8) to (1.9,2);
			\node at (1.75,-0.5) {n$_2$};
			\vertex[dot] (sr) at (1.75,2.5) {};
			\vertex[dot, color=black!0] (b3) at (3.5,1.5) {};
			\draw[color=black!0, ->,>=latex, thick] (3.5,0) to (3.5,3);
			\node[color=black!0] at (3.5,-0.5) {n$_3$};
			\propag[sca, thick] (b1) to [edge label = $\pi^0$](b2);
			\propag[bos, thick] (b2) to [half right, looseness=1.8, edge label' = \vphantom{$\pi^0$}sr] (sr);
		\end{feynhand}
		\end{tikzpicture}
	\end{subfigure}
	\begin{subfigure}{.4\textwidth}
		\centering
		\begin{tikzpicture}
		\begin{feynhand}
			\vertex[dot] (b1) at (0,1.5) {};
			\draw[->,>=latex, thick] (0,0) to (0,3);
			\node at (0,-0.5) {n$_1$};
			\vertex[squaredot] (b2) at (1.75,1.5) {};
			\draw[->,>=latex, thick] (1.75,0) to (1.75,3);
			\draw[-, thick] (1.6,1) to (1.9,1.2);
			\draw[-, thick] (1.6,0.8) to (1.9,1);
			\node at (1.75,-0.5) {n$_2$};
			\vertex[dot] (sr) at (1.75,0.5) {};
			\vertex[dot, color=black!0] (b3) at (3.5,1.5) {};
			\draw[color=black!0, ->,>=latex, thick] (3.5,0) to (3.5,3);
			\node[color=black!0] at (3.5,-0.5) {n$_3$};
			\propag[sca, thick] (b1) to [edge label = $\pi^0$](b2);
			\propag[bos, thick] (b2) to [half left, looseness=1.8, edge label = \vphantom{$\pi^0$}sr] (sr);
		\end{feynhand}
		\end{tikzpicture}
	\end{subfigure}
	\caption{Feynman diagrams from short-range vertex corrections yielding $V_{med}^{(1)}$\,. Graphs arising from the exchange $n_1 \leftrightarrow n_2$ are not shown and have to be added.}
	\label{fig:srvc}
\end{figure}

\begin{figure}[H]
	\centering
	\begin{subfigure}{.4\textwidth}
		\centering
		\begin{tikzpicture}
		\begin{feynhand}
			\vertex[dot, color=black!0] (b1) at (0,1.5) {};
			\vertex[dot] (p) at (1.75,2.5) {};
			\draw[color=black!0, ->,>=latex, thick] (0,0) to (0,3);
			\node[color=black!0] at (0,-0.5) {n$_1$};
			\vertex[squaredot] (b2) at (1.75,1.5) {};
			\draw[->,>=latex, thick] (1.75,0) to (1.75,3);
			\draw[-, thick] (1.6,2) to (1.9,2.2);
			\draw[-, thick] (1.6,1.8) to (1.9,2);
			\node at (1.75,-0.5) {n$_1$};
			\vertex[dot] (b3) at (3.5,1.5) {};
			\draw[->,>=latex, thick] (3.5,0) to (3.5,3);
			\node at (3.5,-0.5) {n$_2$};
			\propag[sca, thick] (b2) to [half left, looseness=1.8, edge label = $\pi^0$](p);
			\propag[bos, thick] (b2) to [edge label = \vphantom{$\pi^0$}sr ] (b3);
		\end{feynhand}
		\end{tikzpicture}
	\end{subfigure}
	\begin{subfigure}{.4\textwidth}
		\centering
		\begin{tikzpicture}
		\begin{feynhand}
			\vertex[dot, color=black!0] (b1) at (0,1.5) {};
			\vertex[dot] (p) at (1.75,0.5) {};
			\draw[color=black!0, ->,>=latex, thick] (0,0) to (0,3);
			\node[color=black!0] at (0,-0.5) {n$_1$};
			\vertex[squaredot] (b2) at (1.75,1.5) {};
			\draw[->,>=latex, thick] (1.75,0) to (1.75,3);
			\draw[-, thick] (1.6,1) to (1.9,1.2);
			\draw[-, thick] (1.6,0.8) to (1.9,1);
			\node at (1.75,-0.5) {n$_1$};
			\vertex[dot] (b3) at (3.5,1.5) {};
			\draw[->,>=latex, thick] (3.5,0) to (3.5,3);
			\node at (3.5,-0.5) {n$_2$};
			\propag[sca, thick] (b2) to [half right, looseness=1.8, edge label' = $\pi^0$](p);
			\propag[bos, thick] (b2) to [edge label = \vphantom{$\pi^0$}sr ] (b3);
		\end{feynhand}
		\end{tikzpicture}
	\end{subfigure}
	\caption{Feynman diagrams from pionic vertex corrections yielding $V_{med}^{(2)}$\,. Graphs arising from the exchange $n_1 \leftrightarrow n_2$ are not shown and have to be added.}
	\label{fig:pvc}
\end{figure}

In these Feynman diagrams, it is possible to see two parallel slashes on the neutron lines within the loop. This denotes the medium insertion of the neutron propagator coming from the filled Fermi sea. Its origin and form are described in the next section. \\\

Having specified the basic mode of 3n-interaction and the topologies arising from closing one neutron line to a loop, the construction of the in-medium nn-potential from subleading chiral 3n-forces is outlined in the next section.

\newpage

\section{Construction of the Effective Neutron-Neutron Potential}\label{sec:method}

In this section, the calculational methods employed in this work are laid out and explained, supplementing background information when it is required. \\\

A basic component of the aforementioned approach to the treatment of 3n-forces is the neutron propagator. As one now considers in-medium interactions, it is required to replace the free neutron propagator by its in-medium counterpart, visible in expression (\ref{eq:phprop}) below. It describes a particle propagating freely outside the Fermi sphere (left term), while a hole propagates within the Fermi sphere (right term) - as detailed on page 26 of ref. \cite{insertion}. Reinstalling a factor $i$ with respect to ref. \cite{insertion}, the in-medium neutron propagator takes on the form

\begin{equation}
	\frac{i\,\theta (\, |\,\vec{l}\,| - k_n\, )}{l^0 - E(\,\vec{l}\,) + i\epsilon} + \frac{i\,\theta (\, k_n - |\,\vec{l}\,| \, )}{l^0 - E(\,\vec{l}\,) - i\epsilon},
	\label{eq:phprop}
\end{equation}

where $\theta(x)$ is the usual Heaviside step function, and $l^0,\, \vec{l}$ are the components of the neutron four-momentum - the neutron energy and its three-momentum. Furthermore, $k_n$ denotes the neutron Fermi momentum, which is related to the neutron density by $\rho_n = k_n^3/3\pi^2$. $E(\,\vec{l}\,) = \vec{l}\,^2/2M$ is the kinetic energy with $M = 940\, \si{MeV}$ the neutron mass \cite{nmass}, and $\epsilon$ is an infinitesimal positive parameter to properly treat the poles.\\
However, by employing the identity

\begin{equation}
	\frac{1}{x \pm i\epsilon} = \mathcal{P}\frac{1}{x} \mp i\pi \delta(x),
\end{equation}

it is possible to derive another representation, which is particularly useful in the following computations. Here, $\mathcal{P}$ stands for the Cauchy principal value, and $\delta(x)$ denotes the Dirac delta-function. Thus, after multiplying with the imaginary unit, the in-medium neutron propagator reads:

\begin{equation}
	\begin{aligned}
		& \frac{i}{l^0 - E(\,\vec{l}\,) + i\epsilon} - 2\pi \delta (\,l^0 - E(\,\vec{l}\,))\, \theta (\,k_n - |\,\vec{l}\,|\,) \\[0.5em]
		& \longrightarrow \frac{i}{ \vphantom{\vec{l}} l^0 + i \epsilon} - 2\pi \delta(\,l^0\,)\, \theta (\,k_n - |\,\vec{l}\,|\,).
	\end{aligned}
	\label{eq:nprop}
\end{equation}

In the second line, the heavy baryon limit $M \rightarrow \infty$ was employed. It is equivalent to the scaling $l^0 \backsimeq |\,\vec{l}\,|$, such that terms of order $\mathcal{O}(\,\vec{l}\,^2\,)$ can be neglected. \\
The first term describes a free propagation and does not yield any $k_n$-dependent contributions, thus being irrelevant henceforth. The crucial second part leads to density-dependent terms when performing the necessary loop integral $(2\pi)^{-4}\int d^4l$, and it is called the in-medium insertion. Making use of the delta- and theta-functions, one finally obtains the Fermi sphere integral present in all calculations to obtain the effective nn-interaction, 

\begin{equation}
	- \int\limits_{|\,\vec{l}\,| \,<\, k_n} \frac{d^3l}{(2\pi)^3}\,.
\end{equation}

After covering the medium insertion as the salient feature for computations, it is now time to consider the kinematical variables and operators constituting the 3n-interactions, and effective nn-potential. \\
Within the explicit form of the 3N-forces taken from ref. \cite{sr}, a number of quantities appear, namely the in- and outgoing particle momenta $\vec{p}_i,\, \vec{p}_i\hspace{-0.07cm}'$\,, as well as the spin- and isospin-vector operators $\vec{\sigma}_i,\, \vec{\tau}_i$\,, where $i = 1, 2, 3$ denotes the respective nucleon. Furthermore, the three-momentum-transfers, defined as $\vec{q}_i = \vec{p}_i\hspace{-0.07cm}' - \vec{p}_i$\,, are encountered frequently. All of the appearing 3N-force terms are expressed by these variables. Also, using conservation of momentum, it is easy to verify that the momentum-transfers satisfy the constraint $\vec{q}_1 + \vec{q}_2 + \vec{q}_3 = 0$.\\
In order to facilitate the simplest possible solution, all calculations are performed in the center-of-mass (CM) frame throughout this work. In practice, this means that with respect to the in- and outgoing neutrons of the effective nn-interaction, the reference frame is chosen such that $\vec{p}_1 = - \vec{p}_2 \equiv \vec{p},\, \vec{p}_1\hspace{-0.07cm}' = - \vec{p}_2\hspace{-0.07cm}' \equiv \vec{p}\,'$ holds. \\
Additionally, elastic on-shell scattering is assumed, meaning that $|\,\vec{p}\,| = |\,\vec{p}\,'\,| \equiv p$, leading to the restraints on the momentum-transfer modulus $0 \leq q \leq 2p$, where $q \equiv |\,\vec{q}\,| = 2p \sin (\theta / 2) = p \sqrt{2(1-z)}$ with $z= \cos \theta$ and $\theta$ the CM scattering angle.\\

Having defined the kinematical quantities, it is now possible to understand the various operators appearing in the effective nn-potential, constructed from spin-vector operators and momenta of the involved two neutrons. \\
Beginning with the spin-operators, there are 5 distinct terms, which do not change within the medium when compared to their counterparts in free space,

\begin{equation}
	1, \hspace{0.5cm} \vec{\sigma}_1 \cdot \vec{\sigma}_2\,, \hspace{0.5cm} \vec{\sigma}_1 \cdot \vec{q}\, \vec{\sigma}_2 \cdot \vec{q}\,, \hspace{0.5cm} i\, (\vec{\sigma}_1 + \vec{\sigma}_2) \cdot (\vec{q} \cross \vec{p}), \hspace{0.5cm} \vec{\sigma}_1 \cdot \vec{p}\, \vec{\sigma}_2 \cdot \vec{p} + \vec{\sigma}_1 \cdot \vec{p}\,'\, \vec{\sigma}_2 \cdot \vec{p}\,'.
\end{equation}

In the given order, the first four are called central, spin-spin, tensor and spin-orbit terms. The last term is related to the quadratic spin-orbit operator by the following identity:

\begin{equation}
	\begin{aligned}
		\vec{\sigma}_1 \cdot (\vec{q} \cross \vec{p})\, \vec{\sigma}_2 \cdot (\vec{q} \cross \vec{p}) =\, & q^2 \left(p^2 - \frac{q^2}{4} \right) \vec{\sigma}_1 \cdot \vec{\sigma}_2 + \left( \frac{q^2}{2} - p^2 \right) \vec{\sigma}_1 \cdot \vec{q}\, \vec{\sigma}_2 \cdot \vec{q} \\[0.5em]
		& - \frac{q^2}{2} (\vec{\sigma}_1 \cdot \vec{p}\, \vec{\sigma}_2 \cdot \vec{p} + \vec{\sigma}_1 \cdot \vec{p}\,'\, \vec{\sigma}_2 \cdot \vec{p}\,').
		\end{aligned}
	\label{eq:qso}
\end{equation}

This decomposition can be checked by making use of the Levi-Civita symbol identity:

\begin{equation}
 \epsilon_{ijk}\epsilon_{lmn} = 
 \begin{vmatrix} 
 \delta_{il} & \delta_{im} & \delta_{in} \\
 \delta_{jl} & \delta_{jm} & \delta_{jn} \\
 \delta_{kl} & \delta_{km} & \delta_{kn}
 \end{vmatrix},
\end{equation}

where $\epsilon_{ijk}$ and $\delta_{ij}$ denote the Levi-Civita symbol and the Kronecker delta, respectively. \\
As a side-note, many of the calculations require the application of the following useful identities for products of Pauli spin matrices:

\begin{equation}
	\vec{\sigma} \cdot \vec{a}\, \vec{\sigma} \cdot \vec{b}  = \vec{a} \cdot \vec{b} + i\, \vec{\sigma} \cdot (\,\vec{a} \cross \vec{b}\,), 
\end{equation}
\vspace{-0.35cm}
\begin{equation}
	\epsilon_{ijk}\, \sigma_j\, \sigma_k  = 2 i\, \sigma_i\,,
\end{equation}

where $\vec{a},\, \vec{b}$ are arbitrary three-vectors. \\\

Concerning the occurring isospin-vector operators within the in-medium NN-potential in isospin-symmetric nuclear matter, these only encompass the two possible structures

\begin{equation}
	1 \hspace{0.75cm} \text{and} \hspace{0.75cm} \vec{\tau}_1 \cdot \vec{\tau}_2\,,
\end{equation}

which are identical to the ones for the NN-potential in free space. \\
This leads directly into the key difference between isospin-symmetric matter, involving equal parts protons and neutrons, and pure neutron matter. That is, the substitutions $\vec{\tau}_i \cdot \vec{\tau}_j \rightarrow 1,\, i,j = 1, 2, 3$ and $\vec{\tau}_1 \cdot (\vec{\tau}_2 \cross \vec{\tau}_3) \rightarrow 0$ hold. \\
In order to derive these conditions, one has to remember that the operators for three neutrons act on the neutron state in isospin-space. Thus, by using the isospin expectation value in pure neutron matter,

\begin{equation}
	\bra{n} \vec{\tau}_i \ket{n} = \begin{pmatrix} 0 \\ 0 \\ -1 \end{pmatrix},
\end{equation}

one obtains

\begin{equation}
	\bra{n\,n} \vec{\tau}_i \cdot \vec{\tau}_j \ket{n\,n} = \begin{pmatrix} 0 \\ 0 \\ -1 \end{pmatrix} \cdot \begin{pmatrix} 0 \\ 0 \\ -1 \end{pmatrix} = 1
\end{equation}

and

\begin{equation}
	\bra{n\,n} \vec{\tau}_1 \cdot (\vec{\tau}_2 \cross \vec{\tau}_3) \ket{n\,n} = \begin{pmatrix} 0 \\ 0 \\ -1 \end{pmatrix} \cdot \left[ \begin{pmatrix} 0 \\ 0 \\ -1 \end{pmatrix} \cross \begin{pmatrix} 0 \\ 0 \\ -1 \end{pmatrix} \right] = 0\,.
\end{equation}

Hence, by plugging in the aforementioned substitutions into the subleading 3N-forces, one obtains easily the corresponding 3n-forces. With respect to isospin-symmetric matter, this leads to many contributions to the in-medium nn-potential that differ only by a constant isospin factor, whereas others - especially those involving the isospin-vector triple product - may vary greatly. In particular for self-closings, the changed isospin operators give rise to non-zero contributions in pure neutron matter, while the counterparts in the effective NN-interaction in symmetric nuclear matter often vanish. The reason for this will become apparent later during this chapter, and is discussed in detail in the following chapters \ref{chapter:results} and \ref{chapter:conclusion}, as well. \\\

Before moving on to the explicit method of calculating Fermi sphere integrals, additional pieces of relevant information have to be given. As such, since the presented calculations are performed under the assumption of spin-saturated neutron matter as well and thus containing the same number of spin-up and -down particles, the aforementioned neutron density is given by $\rho_n = k_n^3/3\pi^2$. Notice the difference to the nucleon density in isospin-symmetric matter, $\rho = 2k_n^3/3\pi^2$, due to a missing isospin-multiplicity factor of $2$ for the two Fermi seas of protons or neutrons. \\
Lastly, throughout this work, the same sign-convention is followed as in ref. \cite{sr}, such that at tree-level, the one-pion-exchange nn-potential is given by $V_{1\pi} = - (g_A/2f_\pi)^2 \\ \times (m_\pi^2 + q^2)^{-1}\, \vec{\sigma}_1 \cdot \vec{q}\, \vec{\sigma}_2 \cdot \vec{q}$\,. Here, $g_A = 1.3$ denotes the axial-vector coupling constant of the nucleon \cite{LO, forces}, $f_\pi = 92.2\, \si{MeV}$ is the weak pion-decay constant \cite{LO, forces}, and $m_\pi = 135\, \si{MeV}$ the neutral pion mass \cite{mpi}. In contrast to ref. \cite{sr}, $m_\pi$ is set to be the neutral pion mass instead of the average mass among the pion triplet, since the charged pions are not involved, as previously explained. \\\

After covering the basic ingredients needed in this work, it is now time to move on to the calculation of Fermi sphere integrals.\\
During the derivation of effective nn-potentials, whose results are presented in the following chapter, one encounters Fermi sphere integrals over even functions \\ $F(s) = F(-s)$, where $s^2 = (\, \vec{l} + \vec{p}\, )^2 = p^2 + l^2 + 2lpz$\,, and $z = \cos\alpha$ with $\alpha$ being the angle between $\vec{l}$ and $\vec{p}$\,. Making use of $F(s)$ being even, these can be reduced to one-parameter integrals in the following way:

\begin{equation}
	\begin{aligned}
		\int\limits_{|\,\vec{l}\,| \,<\, k_n} \frac{d^3l}{2\pi}\, F(s) & = \int\limits_0^{k_n} dl\, l^2 \int\limits_{-1}^{1} dz \,F(s) =  \int\limits_0^{k_n}dl\, \frac{l}{p} \int\limits_{|p-l|}^{p+l} ds\, s\, F(s) \\[0.5em]
		& = \int\limits_0^{k_n}dl\, \frac{l}{p} \int\limits_{p-l}^{p+l} ds\, s\, F(s) =  \int\limits_{p-k_n}^{p+k_n} ds\, s\, F(s) \int\limits_{|s-p|}^{k_n} dl\, \frac{l}{p} \\[0.5em]
		& = \int\limits_{p-k_n}^{p+k_n} ds\, \frac{s}{2p} \big[k_n^2 - (s-p)^2 \big] F(s).
	\end{aligned}
\end{equation}

The crucial step consists of dropping the absolute magnitude at $|p-l|$, which is allowed since the antiderivative of $s\,F(s)$ is an even function.\\
Furthermore, Fermi sphere integrals involving tensorial factors $\{l_i,\, l_il_j,\, l_il_jl_k\}$ appear frequently as well. These are solved by making use of the symmetry regarding the exchange of indices, and constructing the appropriate general form of the integral in terms of tensorial factors built from $\delta_{ij}$ and $p_i$ on the one hand, and scalar loop-functions on the other hand. Finally, one contracts the integral with the prefactors of the loop-functions, thus obtaining linear equations. Adhering to this method, one can find the following reduction formulae for integrals:

\begin{equation}
	\begin{aligned}
		\int\limits_{|\,\vec{l}\,| \,<\, k_n} \frac{d^3l}{2\pi}\, F(s) \{1,\, l_i,\, l_il_j\} = \int\limits_{p-k_n}^{p+k_n} ds\, & \frac{s}{2p} \big[k_n^2 - (s-p)^2 \big] F(s) \\
		& \times \{1,\, \chi_1\,p_i,\, \chi_2\,\delta_{ij} + \chi_3\,p_ip_j\},
	\end{aligned}
	\label{eq:integral}
\end{equation}

which are employed throughout this work, and are equally applicable to $\vec{p} \rightarrow \vec{p}\,'$. The weighting functions have been determined through the previously outlined method and read:

\begin{equation}
	\begin{aligned}
		& \chi_1 = \frac{1}{4p^2}(s^2 + 2sp - 3p^2 - k_n^2), \\[0.5em]
		& \chi_2 = \frac{1}{24p^2}\big[k_n^2 - (s-p)^2\big](s^2 + 4sp + p^2 - k_n^2),\\[0.5em]
		& \chi_3 = \frac{1}{8p^4} \big[ k_n^4 + 2k_n^2(p^2-sp-s^2) + (s-p)^2(s^2+4sp+5p^2) \big].
	\end{aligned}
\end{equation}

Using these expressions, and plugging in a pion propagator $\big[m_\pi^2 + (\, \vec{l} + \vec{p}\,)^2 \big]^{-1}$ for $F(s)$, one can obtain the analytical expressions for the employed loop-functions $\Gamma_\nu (p,\,k_n)$,\, $\gamma_\nu (p,\,k_n)$,\, $\nu = 1, \dots, 5$\,, which are given in the appendix. In the case of the tensorial factor $l_il_jl_k$\, $(\nu = 4, 5)$, the loop-functions are obtained by constructing projection operators out of their prefactors, and contracting them with the Fermi sphere integral. Thus, these loop-functions are calculated directly, without first computing the polynomial weighting functions; the underlying methodology is of course identical. \\
Apart from these loop-functions, two other kinds are used as well, $G_\nu (p,\,q,\,k_n)$ and $K_\nu (p,\,q,\,k_n)$,\, $\nu = 1, 2, 3$. Their analytical construction is limited to the reduction to a radial integral by using the Feynman parametrization,

\begin{equation}
\frac{1}{AB} = \int\limits_0^1 dx \frac{1}{\big[ xA + (1-x)B \big]^2}\,,
\end{equation}

where $A, B$ each denote different pion propagators. One executes the angular- and $x$-integrals to obtain a purely radial integral, and lastly solves a system of linear equations. Although they are not analytical, the functions $G_\nu$ and $K_\nu$ are given in a form which requires just a one-parameter numerical integration. \\
For the sake of readability and notational simplicity, the arguments of the loop-functions will be suppressed henceforth.\\\

Armed with the basic tools to evaluate loop integrals and the knowledge of the appearing interaction topologies, it is now possible to understand the procedure of the necessary calculations, as well as the underlying physics of the approach employed in the derivation of the effective in-medium nn-potential. \\
First, the calculations for self-closings are illustrated in more detail. Performing the Fermi sphere integral arising from closing the neutron-loop and making use of the in-medium insertion, it is important to remember the Feynman rules for fermionic loops, leading to an additional factor of minus one. Also, it is necessary to take the trace over the respective neutron spin, meaning that contributions containing the spin-vectors vanish, as $\vec{\sigma}$ is traceless. \\
Terms involving the respective momentum-transfer yield zero as well, due to momentum conservation at every vertex demanding $\vec{q}_i = 0$, where the index $i$ refers to the closed neutron line. Thus, using the previously mentioned relation $\vec{q}_1 + \vec{q}_2 + \vec{q}_3 = 0$, one finds that in this case $\vec{q}_j = - \vec{q_k}$\,, the indices $j, k$ standing for the external neutrons. Hence, as there are no $\vec{l}$-dependent terms, the Fermi sphere integral yields a factor of $\rho_n/2$. In the end, one has to add the individual contributions - though, in most cases only one of the neutron loops leads to a non-zero term - and make the appropriate adjustments to the indices, $(2,3) \rightarrow (1,2)$ when closing $n_1$ or $3 \rightarrow 2$ when closing $n_2$. As final step, the result arising from the mirror diagram is added to the previously determined one.\\\

With respect to each other, the three remaining contributions exhibit very similar approaches to their calculation. \\
One begins by assigning the appropriate momenta $\pm \vec{p}\,,\, \pm \vec{p}\,',\, \pm \vec{l}$ to the $\vec{p}_i\,,\, \vec{p}_i\,'$ and thus derive the expressions for the corresponding momentum-transfers $\vec{q}_i$\,. To the end of simplifying calculations, it is advantageous to choose $+\vec{l}$ as three-momentum at the medium insertion for short-range vertex corrections, and $-\vec{l}$ for pionic vertex corrections and double exchanges. As the Fermi sphere is invariant under $\vec{l} \leftrightarrow -\vec{l}$\,, one has the freedom to choose the more convenient option.\\ 
After doing this for both types of Feynman diagrams with a set constellation of $n_1$ and $n_2$, it is advisable to check that the derived $\vec{q}_i$ satisfy $\vec{q}_1 + \vec{q}_2 + \vec{q}_3 = 0$. Subsequently, one writes down the Fermi sphere integrals for the given interaction, plugging in the derived expressions above, and taking care to arrange the spin operators corresponding to the order they are applied to the neutron line. Then, one reassigns the neutron indices, namely $(2,3) \rightarrow (1,2)$ for pionic vertex corrections and $3 \rightarrow 1$ for double exchanges. Finally, it is most convenient and efficient to omit the contributions arising from the exchange $n_1 \leftrightarrow n_2$ from the initial calculations, and only add them to $V_{med}$ in the end. \\\

Thus, after explaining the employed tools and necessary information, it is now possible to present the resulting in-medium nn-potentials in the following chapter, together with the underlying calculations to obtain them.

\newpage

\chapter{Resulting Contributions to the In-Medium Neutron-Neutron Interaction}\label{chapter:results}

In this chapter, the results for contributions to the in-medium nn-potential $V_{med}$\,, obtained by closing one neutron line of the subleading chiral 3n-forces, are given explicitly. They are expressed in terms of the loop-functions defined in the appendix, and discussed subsequently. \\ First, the method is applied to the short-range one- and two-pion-exchange-contact topologies in sections (\ref{sec:1pec}) and (\ref{sec:2pec}) respectively. After that, the contributions arising from relativistic corrections are presented in section (\ref{sec:rel}), divided further into a one-pion-exchange-contact topology, as well as a two-pion-exchange topology.

\section{One-Pion-Exchange-Contact Topology}\label{sec:1pec}

Starting with the $1\pi$-exchange-contact topology, there are two contributions to the 3n-interaction:

\begin{equation}
	V_{3n} \ \ = \ \ - \frac{g_A^4 C_T m_\pi}{16\pi f_\pi^4} \frac{\vec{\sigma}_1 \cdot \vec{q}_1\, \vec{\sigma}_3 \cdot \vec{q}_1}{m_\pi^2 + q_1^2}
\end{equation}

and

\begin{equation}
	V_{3n} \ \ = \ \ \frac{g_A^4 C_T m_\pi}{16\pi f_\pi^4} \frac{\vec{\sigma}_1 \cdot \vec{q}_1\, \vec{\sigma}_3 \cdot \vec{q}_1}{m_\pi^2 + q_1^2}\,,
\end{equation}

taken from eqs. (3) and (4) of ref. \cite{sr}, respectively. The operator substitution is performed by making use of both isospin identities mentioned in the previous chapter. \\
In the interactions above, $C_T$ denotes a low-energy constant assigned to the leading spin-dependent NN-contact interaction \cite{forces}. \\
As the terms are opposite in sign and identical otherwise, they add up to zero, yielding no net contribution within the $1\pi$-exchange-contact topology. This is consistent with ref. \cite{forces}, where these 3N-forces vanish by total antisymmetrization, and ref. \cite{sr} in which the contributions to the NN-potential in isospin-symmetric matter cancel within each partial wave.\\\

This section on the $1\pi$-exchange-contact topology is followed by the contributions arising from the $2\pi$-exchange-contact topology interactions, which are discussed on the following pages.

\newpage

\section{Two-Pion-Exchange-Contact Topology}\label{sec:2pec}

Within the $2\pi$-exchange-contact topology, two contributions have been derived in ref. \cite{forces} for the 3N-force. The first one, adapted to the 3n-case, reads:

\begin{equation}
	V_{3n} \ \ = \ \ - \frac{g_A^2 C_T}{24\pi f_\pi^4}\, \vec{\sigma}_2 \cdot \vec{\sigma}_3 \Big[m_\pi + (2m_\pi^2 + q_1^2)A(q_1)\Big],
\end{equation}

where

\begin{equation}
	A(q_1) \ \ = \ \ \frac{1}{2q_1} \arctan \frac{q_1}{2m_\pi}
\end{equation}

the relevant pion-loop function. These expressions are taken from eqs. (8) and (9) of ref. \cite{sr}, after making use of the identity $\vec{\tau}_1 \cdot \vec{\tau}_2 = 1$ for pure neutron states. \\
Using the limit of zero momentum-transfer imposed by the self-closings as seen in fig. (\ref{fig:sc}), and the thus resulting factor

\begin{equation}
	A(0) \ \ = \ \ \frac{1}{4m_\pi}\,,
\end{equation}

one obtains the contribution

\begin{equation}
	\begin{aligned}
		V_{med}^{(0)} & \ \ = \ \ - \frac{g_A^2 C_T m_\pi k_n^3}{24\pi^3f_\pi^4}\, \vec{\sigma}_1 \cdot \vec{\sigma}_2 \\[0.5em]
		& \ \ = \ \ \frac{g_A^2 C_T m_\pi k_n^3}{8\pi^3f_\pi^4}\,.
	\end{aligned}
	\label{eq:2p0}
\end{equation}

While the analogous term in isospin-symmetric nuclear matter was zero due to a vanishing isospin trace, this is evidently not the case in pure neutron matter. \\
Furthermore, $\vec{\sigma}_1 \cdot \vec{\sigma}_2 = - \vec{\sigma}_1 \cdot \vec{\sigma}_1 = -3$ can be used in the second line of equation (\ref{eq:2p0}), as $V_{med}^{(0)}$ contains only a spin-spin coupling term, and does not depend on $q^2 = 2p^2(1-z)$ and thus the scattering angle in the CM frame, leading to a total spin of $S = 0$. This can be demonstrated by employing the partial wave projection formulae in eqs. (6) - (9) of ref. \cite{part}, and making use of the orthogonality of ordinary Lergendre polynomials. This approach yields the conditions $(LSJ) = (000),\, (101)$, with $L$ denoting the total orbital angular momentum, and $J$ the total angular momentum. The second of the aforementioned options is inapplicable, however, since fulfillment of $L + S + I = odd$ is demanded, in order to account for an antisymmetric total fermion wave-function under exchange of nucleons. In this case, as the total isospin satisfies $I = 1$ for pure neutron matter, the relevant condition is given by $L + S = even$. Hence, using regular spectroscopic notation, eq. (\ref{eq:2p0}) contributes to the $^{2S+1}L_J =\hspace{0.1cm}\hspace{-0.1cm} ^1S_0$ partial wave only, where the spins fulfill $\vec{\sigma}_2 = - \vec{\sigma}_1$\,, effectively leading to an isotropic interaction. \\\

Next, the contribution from short-range vertex corrections - as seen in fig. (\ref{fig:srvc}) - reads:

\begin{equation}\boxed{
	V_{med}^{(1)} \ \ = \ \ \frac{g_A^2 C_T k_n^3}{12\pi^3f_\pi^4} \Big[m_\pi + (2m_\pi^2 + q^2)A(q)\Big],}
\end{equation} 

which is identical to the result in eq. (10) of ref. \cite{sr}, provided the appropriate adaption to neutron matter $\vec{\tau}_1 \cdot \vec{\tau}_2 \rightarrow 1$ is made.  \\\

Adding the in this case identical contributions from pionic vertex corrections and double exchanges, shown in figs. (\ref{fig:pvc}) and (\ref{fig:de}) respectively, one obtains

\begin{equation}
	\begin{aligned}
		V_{med}^{(2)} + V_{med}^{(3)} \ \ = \ \ & \frac{g_A^2 C_T}{360\pi^3 f_\pi^4}\, \vec{\sigma}_1 \cdot \vec{\sigma}_2 \left\{m_\pi k_n (p^2 + 11k_n^2 - 4m_\pi^2)    \vphantom{\frac{k_n^2}{m_\pi^2}} \right. \\[0.5em]
		& + (p + k_n)^2 \left[\frac{k_n}{p}(k_n^2 + 10m_\pi^2) - 5m_\pi^2 + \frac{7k_n^2}{4} + \frac{pk_n}{2} - \frac{p^2}{4}  \right] \arctan \frac{p + k_n}{2m_\pi} \\[0.5em]
		& + (p - k_n)^2 \left[\frac{k_n}{p}(k_n^2 + 10m_\pi^2) + 5m_\pi^2 - \frac{7k_n^2}{4} + \frac{pk_n}{2} + \frac{p^2}{4}  \right] \arctan \frac{p - k_n}{2m_\pi} \\[0.5em]
		& + \left. \frac{m_\pi^3}{p}(5p^2 - 5k_n^2 + 4m_\pi^2) \ln \frac{4m_\pi^2 + (p + k_n)^2}{4m_\pi^2 + (p - k_n)^2} \right\},
	\end{aligned}
	\label{eq:2p23}
\end{equation}

which, following the same argumentation as for eq. (\ref{eq:2p0}), contributes only to the $^1S_0$-wave. \\
Thus, it is possible to add the three contributions acting only in the $^1S_0$ partial-wave to obtain:\\
\begin{equation}\boxed{
	\begin{aligned}
		V_{med}^{(0 + 2 + 3)} \ \ = \ \ & - \frac{g_A^2 C_T}{120\pi^3 f_\pi^4} \left\{m_\pi k_n (p^2 - 4k_n^2 - 4m_\pi^2)    \vphantom{\frac{k_n^2}{m_\pi^2}} \right. \\[0.5em]
		& + (p + k_n)^2 \left[\frac{k_n}{p}(k_n^2 + 10m_\pi^2) - 5m_\pi^2 + \frac{7k_n^2}{4} + \frac{pk_n}{2} - \frac{p^2}{4}  \right] \arctan \frac{p + k_n}{2m_\pi} \\[0.5em]
		& + (p - k_n)^2 \left[\frac{k_n}{p}(k_n^2 + 10m_\pi^2) + 5m_\pi^2 - \frac{7k_n^2}{4} + \frac{pk_n}{2} + \frac{p^2}{4}  \right] \arctan \frac{p - k_n}{2m_\pi} \\[0.5em]
		& + \left. \frac{m_\pi^3}{p}(5p^2 - 5k_n^2 + 4m_\pi^2) \ln \frac{4m_\pi^2 + (p + k_n)^2}{4m_\pi^2 + (p - k_n)^2} \right\},
	\end{aligned}}
	\label{eq:2p023}
\end{equation}

where a new notation was employed for the sake of readability, defining $V_{med}^{(0 + 2 + 3)} \equiv V_{med}^{(0)} + V_{med}^{(2)} + V_{med}^{(3)}\,$.
Additionally, it is interesting to note that eq. (\ref{eq:2p23}) is identical to the result in eq. (11) of ref. \cite{sr}, provided the appropriate substitution $3 + \vec{\tau}_1 \cdot \vec{\tau}_2 \rightarrow 2$ is made, to account for the difference in isospin factors. Although this is a recurring feature throughout this work, it cannot be generally assumed that this holds for all contributions apart from self-closings. As previously seen, due to the product of two isospin vectors always giving 1 for neutrons, this may lead to now non-vanishing terms. This becomes apparent both when calculating the contributions arising from the second interaction term in this topology, as well as the ones given by the leading relativistic corrections in section (\ref{sec:rel}). \\\

The aforementioned second interaction term from the $2\pi$-exchange-contact topology is given in eq. (12) of ref. \cite{sr}, and after adaption to the 3n-case it reads:

\begin{equation}
	\begin{aligned}
		V_{3n} \ \ = \ \ & \frac{g_A^4C_T}{48\pi f_\pi^4} \left\{ 2\, \vec{\sigma}_2 \cdot \vec{\sigma}_3 \left[3m_\pi - \frac{m_\pi^3}{4m_\pi^2 + q_1^2} + 2(2m_\pi^2 + q_1^2)A(q_1) \right] \right. \\[0.5em]
		& + \left. 9 [\vec{\sigma}_1 \cdot \vec{q}_1\, \vec{\sigma}_2 \cdot \vec{q_1} - q_1^2\, \vec{\sigma}_1 \cdot \vec{\sigma}_2]A(q_1)   \vphantom{\frac{k_n^2}{m_\pi^2}} \right\}.
		\end{aligned}
		\label{eq:v3n2p2}
\end{equation}

In this particular case, it is convenient to add the contributions of self-closings from the second and third neutron line (here denoted as $V_{med}^{(0')}\,$) and the short-range vertex corrections, to obtain

\begin{equation}\boxed{
	V_{med}^{(0')} + V_{med}^{(1)} \ \ = \ \ \frac{g_A^4 C_T k_n^3}{12\pi^3 f_\pi^4} \left[ \frac{m_\pi^3}{4m_\pi^2 + q^2} - 3m_\pi - 2(2m_\pi^2 + q^2)A(q) \right],}
\end{equation}

since $V_{med}^{(0')}$ is canceled by a term present in $V_{med}^{(1)}$\,. \\
In the same way as with eq. (\ref{eq:2p023}), it is advantageous to subsequently add the contributions arising from closing the first neutron line to itself (here denoted as $V_{med}^{(0'')}\,$), the pionic vertex corrections, and double exchanges. Making use of the notation employed in eq. (\ref{eq:2p023}), this yields \\

\begin{equation}\boxed{
	\begin{aligned}
		V_{med}^{(0'' + 2 +3)} \ \ = \ \ & \frac{g_A^4 C_T}{12\pi^3 f_\pi^4} \left\{ \frac{m_\pi k_n}{5} \left(p^2 + \frac{9k_n^2}{4} - 19 m_\pi^2\right)    \right. \\[0.5em]
		& + \left[ \frac{k_n^3}{p} \left(2m_\pi^2 + \frac{k_n^2}{5} \right) + 3 \left(2m_\pi^4 + m_\pi^2 k_n^2 + \frac{k_n^4}{4} \right) + pk_n^3 \right. \\[0.5em]
		& + \left. p^2 \left(\frac{k_n^2}{2} - m_\pi^2 \right) - \frac{p^4}{20} \right] \arctan \frac{p + k_n}{2m_\pi} \\[0.5em]
		& + \left[ \frac{k_n^3}{p} \left(2m_\pi^2 + \frac{k_n^2}{5} \right) - 3 \left(2m_\pi^4 + m_\pi^2 k_n^2 + \frac{k_n^4}{4} \right) + pk_n^3 \right. \\[0.5em]
		& + \left. p^2 \left(m_\pi^2 - \frac{k_n^2}{2} \right) + \frac{p^4}{20} \right] \arctan \frac{p - k_n}{2m_\pi} \\[0.5em]
		& + \left. \frac{m_\pi^3}{p} \left[ \frac{7}{4} (p^2 - k_n^2) - \frac{11 m_\pi^2}{5} \right] \ln \frac{4m_\pi^2 + (p + k_n)^2}{4m_\pi^2 + (p - k_n)^2} \right\}.
	\end{aligned}}
	\label{eq:2p0232}
\end{equation}

Here, the fact that the terms involving the factor q in eq. (\ref{eq:v3n2p2}) cancel each other when adding $V_{med}^{(2)}$ and $V_{med}^{(3)}$ was already made use of, leaving a contribution which only contributes to the $^1S_0$ partial-wave. This holds true for $V_{med}^{(0'')}$ as well, hence the addition to the combined contribution in eq. (\ref{eq:2p0232}). \\
The notable contrast in length compared to the result presented in eq. (14) of ref. \cite{sr} stems from the cancellation of terms proportional to $(\,\vec{l} + \vec{p}\,)^2\, A(|\,\vec{l} + \vec{p}\,|) + (\,\vec{l} + \vec{p}\,'\,)^2\\ \times A(|\,\vec{l} + \vec{p}\,'\,|)$ inside the Fermi sphere integrals for $V_{med}^{(2)}$ and $V_{med}^{(3)}$\,. \\\

The integrals appearing in this section have been computed using eq. (\ref{eq:integral}), derived in the previous chapter. \\\

This concludes the section on the $2\pi$-exchange-contact topology. Thus, as the last - albeit the most expansive - entry within the chapter on results of $V_{med}$\,, contributions arising from the leading relativistic corrections to the chiral 3n-interaction are presented in the following section.

\newpage

\section{Leading Relativistic Corrections}\label{sec:rel}

In this section, contributions to the in-medium nn-potential from the leading relativistic correction terms to chiral 3n-forces are presented. These expressions arise from $1/M$-corrections to the leading $\pi$nn and $\pi\pi$nn vertices on the one hand, and retardation effects on the other hand. \\
Moreover, the 3n-force expressions treated in this section are subdivided into the $1\pi$-exchange-contact topology of subsection (\ref{ssec:r1p}) with contributions proportional to $g_A^2C_{T,S}/Mf_\pi^2$\,, and the $2\pi$-exchange topology of subsection (\ref{ssec:r2p}) with contributions proportional to $g_A^4/Mf_\pi^4$\,. Here, $C_S$ denotes another low-energy constant, associated with the lowest-order spin-independent NN-contact interaction \cite{forces}. \\
One should note that the following calculations and expressions are significantly more complex than the previous ones, as they - in most cases - involve Fermi sphere integrals over one or multiple pion propagators, and spin-operator orderings have to be considered before changing neutron indices. 

\subsection{One-Pion-Exchange-Contact Topology}\label{ssec:r1p}

Within this topology, there are two $V_{3N}$ terms, both of which are proportional to $\vec{\tau}_1 \cdot \vec{\tau}_2$ and only consist of terms containing either $\vec{\sigma}_3$ or $\vec{q}_3$\,. In isospin-symmetric nuclear matter, this leads to vanishing contributions from self-closings $(\, V_{med}^{(0)}\,)$ due to \,$\tr \vec{\tau}_{1, 2} = 0$, and \,$\tr \vec{\sigma}_3 = 0$ or $\vec{q}_3 = 0$. \\
Therefore, as the isospin-vector dot product is equal to 1 for neutron matter, self-closings now yield non-zero contributions, while all others are identical to the isospin-symmetric case when making the appropriate adaption of the isospin factor. Specifically, these changes are $\vec{\tau}_1 \cdot \vec{\tau}_2 \rightarrow 1$ for short-range vertex corrections ($\,V_{med}^{(1)}\,$) and double exchanges ($\,V_{med}^{(3)}\,$). The contribution from pionic vertex corrections ($\,V_{med}^{(2)}\,$) are reduced by a factor of 3, due to the dot product of identical isospin-vector operators $\vec{\tau}_1 \cdot \vec{\tau}_1 = 3$ in isospin-symmetric matter that is replaced by a factor 1 in pure neutron matter. \\\
 
The first part of the 3n-interaction, as taken from eq. (23) of ref. \cite{sr}, reads:

\begin{equation}
	\begin{aligned}	
		V_{3n} \ \ = \ \ & - \frac{g_A^2}{16M f_\pi^2} \frac{1}{m_\pi^2 + q_1^2} \Big\{ C_T \Big[ i\, \vec{\sigma}_1 \cdot (\vec{p}_1 + \vec{p}_1\hspace{-0.07cm}')(\vec{\sigma}_2 \cross \vec{\sigma}_3) \cdot \vec{q}_1 + 3\, \vec{\sigma}_1 \cdot \vec{q}_1\, \vec{\sigma}_3 \cdot \vec{q}_3  \\[0.5em]
		& + 3i\, \vec{\sigma}_1 \cdot \vec{q}_1\, (\vec{\sigma}_2 \cross \vec{\sigma}_3) \cdot (\vec{p}_2 + \vec{p}_2\hspace{-0.07cm}') \Big] + 3C_S\, \vec{\sigma}_1 \cdot \vec{q}_1\, \vec{\sigma}_2 \cdot \vec{q}_3 \Big\},
	\end{aligned}
\end{equation}

and arises out of the $1/M$ correction to the $\pi^0$nn vertex ($\pi$n coupling) and 4n-contact vertex (nn-contact interaction). \\
As previously mentioned, and contrary to the 3N-case, the contribution stemming from self-closings (of the middle n-line) does not vanish for pure neutron matter, yielding

\begin{equation}\boxed{
	V_{med}^{(0)} \ \ = \ \ \frac{g_A^2 C_T k_n^3}{8\pi^2 M f_\pi^2} \frac{\vec{\sigma}_1 \cdot \vec{q}\, \vec{\sigma}_2 \cdot \vec{q}}{m_\pi^2 + q^2}\,.}
\end{equation}

On the other hand, short-range vertex corrections give rise to the contribution

\begin{equation}\boxed{
	V_{med}^{(1)} \ \ = \ \ \frac{g_A^2 k_n^3}{16\pi^2 M f_\pi^2} (C_T - C_S) \frac{\vec{\sigma}_1 \cdot \vec{q}\, \vec{\sigma}_2 \cdot \vec{q}}{m_\pi^2 + q^2}\,,}
\end{equation}

which is proportional to the $1\pi$-exchange nn-interaction potential in momentum space \cite{1ppot}, and linear in the neutron density $\rho_n = k_n^3/3\pi^2$. \\
Next, when considering pionic vertex corrections, one can derive

\begin{equation}\boxed{
	\begin{aligned}
		V_{med}^{(2)} \ \ = \ \ & 
		\frac{g_A^2}{8\pi^2 M f_\pi^2} C_T \left\{ \left[ 2p^2 (\Gamma_3 - \Gamma_0) + \frac{3q^2}{4} (\Gamma_0 + \Gamma_1) + 4\Gamma_2 \right] \vec{\sigma}_1 \cdot \vec{\sigma}_2 \right.  \\[0.5em]
		& - \left. \,\frac{3}{2} (\Gamma_0 + \Gamma_1)\, \vec{\sigma}_1 \cdot \vec{q}\, \vec{\sigma}_2 \cdot \vec{q} + (\Gamma_0 - \Gamma_3) (\vec{\sigma}_1 \cdot \vec{p}\, \vec{\sigma}_2 \cdot \vec{p} + \vec{\sigma}_1 \cdot \vec{p}\,'\, \vec{\sigma}_2 \cdot \vec{p}\,') \right\} \\[0.5em]
		& + \frac{3g_A^2}{32\pi^2 M f_\pi^2} (\Gamma_0 + \Gamma_1) \Big[(C_S-C_T)\, i\, (\vec{\sigma}_1 + \vec{\sigma}_2) \cdot (\vec{q} \cross \vec{p}) - C_S\, q^2 \Big],
	\end{aligned}}
\end{equation}

with double exchanges giving rise to

\begin{equation}\boxed{
	\begin{aligned}
		V_{med}^{(3)} \ \ = \ \ & \frac{g_A^2}{16\pi^2 M f_\pi^2}\, \vec{\sigma}_1 \cdot \vec{\sigma}_2 \left\{ C_T \left[ \frac{3q^2}{2} (\Gamma_0 + \Gamma_1) - p^2 (5\Gamma_0 + 6\Gamma_1 + \Gamma_3) \right] - (2C_T + 3C_S) \Gamma_2 \right\} \\[0.5em]
		& + \frac{g_A^2}{32\pi^2 M f_\pi^2} \Big\{ 3(C_S - C_T)(\Gamma_0 + \Gamma_1)\, \vec{\sigma}_1 \cdot \vec{q}\, \vec{\sigma}_2 \cdot \vec{q}   \\[0.5em]
		& + \Big[ C_T (5\Gamma_0 + 6\Gamma_1 + \Gamma_3) - 3C_S (\Gamma_0 + 2\Gamma_1 + \Gamma_3) \Big] (\vec{\sigma}_1 \cdot \vec{p}\, \vec{\sigma}_2 \cdot \vec{p} + \vec{\sigma}_1 \cdot \vec{p}\,'\, \vec{\sigma}_2 \cdot \vec{p}\,') \Big\} \\[0.5em]
		& + \frac{3g_A^2}{16\pi^2 M f_\pi^2} C_T \left\{ (\Gamma_0 + \Gamma_1) \left[ \frac{q^2}{2} - i\,(\vec{\sigma_1} + \vec{\sigma}_2) \cdot (\vec{q} \cross \vec{p}) \right] - 3\Gamma_2 \right. \\[0.5em]
		& - \left. p^2 (\Gamma_0 + 2\Gamma_1 + \Gamma_3)  \vphantom{\frac{q^2}{2}} \right\}.
	\end{aligned}}
\end{equation}

Besides the already known spin-spin $\big(\hspace{-1mm}\sim \vec{\sigma}_1 \cdot \vec{\sigma}_2\,\big)$ and tensor-type $\big(\hspace{-1mm}\sim \vec{\sigma}_1 \cdot \vec{q}\, \vec{\sigma}_2 \cdot \vec{q}\,\,\big)$ components, both contributions contain a spin-orbit term $\big(\hspace{-1mm}\sim i\,(\vec{\sigma_1} + \vec{\sigma}_2) \cdot (\vec{q} \cross \vec{p})\,\big)$ and a part that involves the quadratic spin-orbit operator according to eq. (\ref{eq:qso}), namely one proportional to $\big(\hspace{-1mm}\sim (\vec{\sigma}_1 \cdot \vec{p}\, \vec{\sigma}_2 \cdot\vec{p} + \vec{\sigma}_1 \cdot \vec{p}\,'\, \vec{\sigma}_2 \cdot \vec{p}\,')\,\big)$. \\
The loop-functions $\Gamma_\nu,\, \nu = 1, \dots , 5$ are defined as Fermi sphere integrals over a single pion propagator, and explained in more detail, as well as given explicitly in the appendix. They involve arctangents and logarithms, and depend on the pion mass $m_\pi$\,, in conjunction with the modulus of the CM on-shell neutron momentum $p$\,, and the neutron Fermi momentum $k_n$\,. This holds true for all following loop-functions, as well. \\\

The second interaction within the $1\pi$-exchange-contact topology, arising from the respective retardation corrections,  reads:

\begin{equation}
	\begin{aligned}
		V_{3n} \ \ = \ \ & \frac{g_A^2}{16 M f_\pi^2} \frac{\vec{\sigma}_1 \cdot \vec{q}_1}{(m_\pi^2 + q_1^2)^2} \Big\{\vec{q}_1 \cdot \vec{q}_3 (C_S\, \vec{\sigma}_2 \cdot \vec{q}_1 + C_T\, \vec{\sigma}_3 \cdot \vec{q}_1) \\[0.5em]
		& + iC_T\, (\vec{\sigma}_2 \cross \vec{\sigma}_3) \cdot \vec{q}_1\, (3\vec{p_1} + 3\vec{p}_1\hspace{-0.07cm}' + \vec{p}_2 + \vec{p}_2\hspace{-0.07cm}') \cdot \vec{q}_1  \Big\},
	\end{aligned}
\end{equation}

which visibly includes a squared pion propagator, and is taken from eq. (27) of ref. \cite{sr}. \\
Closing the middle neutron line to itself leads to the non-vanishing contribution

\begin{equation}\boxed{
	V_{med}^{(0)} \ \ = \ \ - \frac{g_A^2 C_T k_n^3\, q^2}{24 \pi^2 M f_\pi^2} \frac{\vec{\sigma}_1 \cdot \vec{q}\, \vec{\sigma}_2 \cdot \vec{q}}{(m_\pi^2 + q^2)^2}\,,}
\end{equation}

while short-range vertex corrections yield 

\begin{equation}\boxed{
	V_{med}^{(1)} \ \ = \ \ \frac{g_A^2 k_n^3\, q^2}{48 \pi^2 M f_\pi^2} (C_S - C_T) \frac{\vec{\sigma}_1 \cdot \vec{q}\, \vec{\sigma}_2 \cdot \vec{q}}{(m_\pi^2 + q^2)^2}\,,}
\end{equation}

both of which are linear in the neutron density $\rho_n$\,. \\
Computing the results for pionic vertex corrections and double exchanges, one obtains the following comparably long expressions:\\
\begin{equation}\boxed{
	\begin{aligned}
		V_{med}^{(2)} \ \ = \ \ & \frac{g_A^2 C_S\, q^2}{32 \pi^2 M f_\pi^2} \Big[ \Gamma_0 + \Gamma_1 - m_\pi^2 (\gamma_0 + \gamma_1) \Big] \\[0.5em]
		& + \frac{g_A^2 C_T}{8\pi^2 M f_\pi^2} \left\{ (\gamma_2 + \gamma_4)\, \vec{\sigma}_1 \cdot \vec{q}\, \vec{\sigma}_2 \cdot \vec{q} + \left[ 2\Gamma_2 - \frac{4k_n^3}{3} + 2m_\pi^2 (2\Gamma_0 - m_\pi^2\gamma_0 - \gamma_2)	\right. \right. \\[0.5em]
		& + \left. \frac{q^2}{4}(\gamma_2 + \gamma_4) + \left(\frac{q^2}{4} - 4p^2 \right) \big( m_\pi^2 (\gamma_0 + \gamma_1) + \gamma_2 + \gamma_4 - \Gamma_0 - \Gamma_1 \big) \right] \vec{\sigma}_1 \cdot \vec{\sigma}_2 \\[0.5em]
		& + \left[ \Gamma_0 + 2\Gamma_1 + \Gamma_3 - m_\pi^2 (\gamma_0 + 2\gamma_1 + \gamma_3) - 4(\gamma_2 + \gamma_4)  \vphantom{\frac{q^2}{4}} \right. \\[0.5em]
		& + \left. \left. \left(\frac{q^2}{4} - 2p^2 \right) (\gamma_0 + 3\gamma_1 + 3\gamma_3 + \gamma_5) \right] (\vec{\sigma}_1 \cdot \vec{p}\, \vec{\sigma}_2 \cdot \vec{p} + \vec{\sigma}_1 \cdot \vec{p}\,'\, \vec{\sigma}_2 \cdot \vec{p}\,') \right\},
	\end{aligned}}
	\label{eq:typo1}
\end{equation}

and

\begin{equation}\boxed{
	\begin{aligned}
		V_{med}^{(3)} \ \ = \ \ & \frac{g_A^2}{16\pi^2 M f_\pi^2} (C_T - C_S) \big( \gamma_2 + \gamma_4 \big)\, \vec{\sigma}_1 \cdot \vec{q}\, \vec{\sigma}_2 \cdot \vec{q} \\[0.5em]
		& + \frac{g_A^2 C_S}{32\pi^2 M f_\pi^2} \left\{ \Big[ 2\Gamma_2 - (2m_\pi^2 + q^2) \gamma_2 - q^2 \gamma_4 \Big] \vec{\sigma}_1 \cdot \vec{\sigma}_2 + \left[ \Gamma_0 + 2\Gamma_1 + \Gamma_3  \vphantom{\frac{q^2}{2}}   \right. \right. \\[0.5em]
		& - \left. m_\pi^2 (\gamma_0 + 2\gamma_1 + \gamma_3) - \frac{q^2}{2} (\gamma_0 + 3\gamma_1 + 3\gamma_3 + \gamma_5) \right] \\[0.5em]
		& \times \left. \vphantom{\frac{q^2}{2}}  (\vec{\sigma}_1 \cdot \vec{p}\, \vec{\sigma}_2 \cdot \vec{p} + \vec{\sigma}_1 \cdot \vec{p}\,'\, \vec{\sigma}_2 \cdot \vec{p}\,') \right\} \\[0.5em]
		& + \frac{g_A^2 C_T}{16\pi^2 M f_\pi^2} \left\{ \frac{2k_n^3}{3} + m_\pi^2 (m_\pi^2 \gamma_0 - 2\Gamma_0) + \frac{q^2}{2} \Big[m_\pi^2 (\gamma_0 + \gamma_1) - \Gamma_0 - \Gamma_1 \Big] \right. \\[0.5em]
		& + \left[ \left( 4p^2 + \frac{q^2}{2} \right) \Big( m_\pi^2 (\gamma_0 + \gamma_1) + \gamma_2 + \gamma_4 - \Gamma_0 - \Gamma_1 \Big) \right. \\[0.5em]
		& + \left. \vphantom{\frac{q^2}{2}}  2k_n^3 - 3\Gamma_2 + 3m_\pi^2 (\gamma_2 + m_\pi^2 \gamma_0 - 2\Gamma_0) \right] \vec{\sigma}_1 \cdot \vec{\sigma}_2 \\[0.5em]
		& + \left[ \frac{3}{2} \Big( m_\pi^2 (\gamma_0 + 2\gamma_1 + \gamma_3) - \Gamma_0 - 2\Gamma_1 - \Gamma_3 \Big) + 4(\gamma_2 + \gamma_4) \right. \\[0.5em]
		& + \left. \left. \left( 2p^2 + \frac{q^2}{4} \right) (\gamma_0 + 3\gamma_1 + 3\gamma_3 + \gamma_5) \right] (\vec{\sigma}_1 \cdot \vec{p}\, \vec{\sigma}_2 \cdot \vec{p} + \vec{\sigma}_1 \cdot \vec{p}\,'\, \vec{\sigma}_2 \cdot \vec{p}\,') \right\}.
	\end{aligned}}
	\label{eq:typo2}
\end{equation}

The present calculation has revealed typing errors $q^2/8 \rightarrow q^2/4$ in the last line of eq. (29) and $- \Gamma_1 \rightarrow - \Gamma_0 - \Gamma_1$ in the fourth line of eq. (30) in ref. \cite{sr}, which are corrected in the expressions (\ref{eq:typo1}) and (\ref{eq:typo2}), respectively. \\
Furthermore, new loop-functions $\gamma_\nu,\, \nu = 1, \dots , 5$ - defined by Fermi sphere integrals over a squared pion propagator - are introduced, and the corresponding analytical expressions can be found in the appendix.\\\

This concludes the subsection on relativistic corrections to the subleading chiral 3n-force in the $1\pi$-exchange-contact topology. Subsequently, the $2\pi$-exchange topology is discussed in the next subsection.

\newpage

\subsection{Two-Pion-Exchange Topology}\label{ssec:r2p}

The corrections arising from the $2\pi$-exchange topology detailed in this subsection are yet more computationally demanding, as different pion propagators are involved within the same three-neutron interaction $V_{3n}$\,. \\
However, of the four contributions seen in eqs. (31), (34), (37) and (41) of ref. \cite{sr}, only the latter two remain to be considered, as the former two are proportional to the isospin-vector scalar triple product $\vec{\tau}_1 \cdot (\vec{\tau}_2 \cross \vec{\tau}_3)$, which vanishes in the neutron-only case. Thus, there are no contributions to the in-medium nn-interaction from $1/M$- and retardation corrections to a $2\pi$-exchange through two $\pi^0$nn vertices ($\pi$n couplings) and a $\pi\pi$nn (Weinberg-Tomozawa) vertex.\\
As the aforementioned isospin-vector scalar triple product is also present within the two surviving terms, the contributions arising from the corresponding 3n-interactions are noticeably different to the isospin-symmetric case. Nevertheless, when considering double exchanges, the results overlap clearly, as the contributions from the scalar triple product lead to a prefactor of $\vec{\tau}_1 \cdot \vec{\tau}_2$ in symmetric nuclear matter, and the remaining terms differ from the 3n-result only by the previously explained factor of 3 due to a scalar product of two identical isospin vectors. \\
Also, the results for self-closings differ merely by a factor of $1/2$ from those in isospin-symmetric matter. In the latter case, the nucleon loop involves two realizations of the third component of isospin, namely protons and neutrons. \\
Additionally, it is important to keep in mind that in ref. \cite{sr}, both of the pionic vertex corrections have been added to obtain the result for $V_{med}^{(1)}$\,, as it was particularly convenient for the now vanishing interaction terms. This was due to their symmetry under the nucleon exchange $1 \leftrightarrow 3$, yielding the same result for either of the vertex corrections. However, since this symmetry is not applicable to the remaining two 3n-interaction terms, this reduced notation is not employed in this work. \\\

Beginning the presentation of results with the 3N-interaction given in eq. (37) of ref. \cite{sr}, arising from the $1/M$-corrections to the $\pi$NN vertices, the relevant term adapted to the three-neutron case reads:

\begin{equation}
	\begin{aligned}
		V_{3n} \ \ = \ \ & \frac{g_A^4}{64 M f_\pi^4} \frac{\vec{\sigma}_1 \cdot \vec{q}_1}{(m_\pi^2 + q_1^2)(m_\pi^2 + q_3^2)} \left\{ 3\, \vec{\sigma}_3 \cdot \vec{q}_3\, \Big[ i\, \vec{\sigma}_2 \cdot \big( \vec{q}_1 \cross (\vec{p}_2 + \vec{p}_2\hspace{-0.07cm}') \big) + q_1^2 \Big]    \vphantom{\frac{k_n^2}{m_\pi^2}} \right. \\[0.5em]
		& + \left. \vphantom{\frac{k_n^2}{m_\pi^2}} i\, \vec{\sigma}_3 \cdot (\vec{p}_3 + \vec{p}_3\hspace{-0.07cm}')\, \vec{\sigma}_2 \cdot (\vec{q}_1 \cross \vec{q}_3) \right\}.
	\end{aligned}
\end{equation}

This leads to the following contribution from self-closing the middle neutron line:

\begin{equation}\boxed{
	V_{med}^{(0)} \ \ = \ \ - \frac{g_A^4 k_n^3 q^2}{32\pi^2 M f_\pi^4} \frac{\vec{\sigma}_1 \cdot \vec{q}\, \vec{\sigma}_2 \cdot \vec{q}}{(m_\pi^2 + q^2)^2}\,,}
\end{equation}

which - as was previously the case as well - is linear in the neutron density $\rho_n = k_n^3/3\pi^2$.\\
Calculating both types of pionic vertex corrections, one obtains the following contributions to the in-medium nn-potential:

\begin{equation}\boxed{
	V_{med}^{(1)} \ \ = \ \ \frac{g_A^4}{128 \pi^2 M f_\pi^4} \frac{\vec{\sigma}_1 \cdot \vec{q}\, \vec{\sigma}_2 \cdot \vec{q}}{m_\pi^2 + q^2} \Big[ 8p^2 (\Gamma_0 - \Gamma_3) + q^2 (\Gamma_0 + 3\Gamma_1 + 2\Gamma_3) - 16\Gamma_2 \Big],}
\end{equation}

and

\begin{equation}\boxed{
	\begin{aligned}
		V_{med}^{(2)} \ \ = \ \ & \frac{g_A^4}{128 \pi^2 M f_\pi^4} \frac{1}{m_\pi^2 + q^2} \left\{ \Big[ 2k_n^3 + (q^2 - 6p^2) (\Gamma_0 + 2\Gamma_1 + \Gamma_3) - 3m_\pi^2 (\Gamma_0 + \Gamma_1) \Big]    \vphantom{\frac{k_n^2}{m_\pi^2}} \right. \\[0.5em]
		& \times \left.   \vphantom{\frac{k_n^2}{m_\pi^2}} \vec{\sigma}_1 \cdot \vec{q}\,\vec{\sigma}_2 \cdot \vec{q}   + q^2 (\Gamma_0 + 2\Gamma_1 + \Gamma_3) \big(\vec{\sigma}_1 \cdot \vec{p}\, \vec{\sigma}_2 \cdot \vec{p} + \vec{\sigma}_1 \cdot \vec{p}\,'\, \vec{\sigma}_2 \cdot \vec{p}\,'\big) \right\}.
	\end{aligned}}
\end{equation}

Finally, the evaluation of the in-medium loop in the diagram representing double exchanges yields

\begin{equation}\boxed{
	\begin{aligned}
		V_{med}^{(3)} \ \ = \ \ & \frac{g_A^4}{128\pi^2 M f_\pi^4} \left\{ \vphantom{\frac{k_n^2}{m_\pi^2}}    4k_n^3 - 3q^2(\Gamma_0 + \Gamma_1) + 3m_\pi^2 \Big[ (2m_\pi^2 + q^2) G_0 - 4\Gamma_0 \Big]  \right. \\[0.5em]
		& + \Big[ 2G_2 - \Gamma_0 + 2\Gamma_1 + (m_\pi^2 + 2p^2 + q^2) G_0 + 2(4p^2 - q^2) \big(G_1 + G_3 \big) \\[0.5em]
		& - \left. \vphantom{\frac{k_n^2}{m_\pi^2}} 4m_\pi^2 G_1 \Big] i\,(\vec{\sigma}_1 + \vec{\sigma}_2) \cdot (\vec{q} \cross \vec{p}) + 4 \big( G_0 + 2G_1 \big)\, \vec{\sigma}_1 \cdot (\vec{q} \cross \vec{p})\, \vec{\sigma}_2 \cdot (\vec{q} \cross \vec{p}) \right\},
	\end{aligned}}
\end{equation}

which now explicitly includes a term proportional to the quadratic spin-orbit operator $\vec{\sigma}_1 \cdot (\vec{q} \cross \vec{p})\, \vec{\sigma}_2 \cdot (\vec{q} \cross \vec{p})$. Here, it appears directly in the underlying computation, whereas it was previously only present through the operator $(\vec{\sigma}_1 \cdot \vec{p}\, \vec{\sigma}_2 \cdot \vec{p} + \vec{\sigma}_1 \cdot \vec{p}\,'\, \vec{\sigma}_2 \cdot \vec{p}\,')$ after further decomposition. \\
As with the previous loop-functions, the new  $G_\nu,\, \nu = 0, \dots , 3$ - given by Fermi sphere integrals over two different pion propagators - are detailed in the appendix, and exhibit an additional dependence on the momentum-transfer modulus $q$\,. However, unlike the loop-functions introduced up to this point, the $G_\nu$ cannot be given in closed analytical form, but involve a one-parameter integral that needs to be solved numerically.
\\\

The last 3n-interaction term covered in this work arises from retardation corrections to the consecutive $2\pi$-exchange and it is given by

\begin{equation}
	\begin{aligned}
		V_{3n} \ \ = \ \ & \frac{g_A^4}{64M f_\pi^4} \frac{\vec{\sigma}_1 \cdot \vec{q}_1\, \vec{\sigma}_3 \cdot \vec{q}_3}{(m_\pi^2 + q_1^2)^2(m_\pi^2 + q_3^2)} \Big\{ -(\vec{q}_1 \cdot \vec{q}_3)^2 \\[0.5em]
		& + i\, \vec{\sigma}_2 \cdot (\vec{q}_1 \cross \vec{q}_3) (3 \vec{p}_1 + 3 \vec{p}_1\hspace{-0.07cm}' + \vec{p}_2 + \vec{p}_2\hspace{-0.07cm}') \cdot \vec{q}_1 \Big\},
	\end{aligned}
	\label{eq:r2p2}
\end{equation}

where eq. (41) of ref. \cite{sr} was adapted to the three-neutron case. \\
Following the same methods as before, the contribution

\begin{equation}\boxed{
	V_{med}^{(0)} \ \ = \ \ \frac{g_A^4 k_n^3\, q^4}{96\pi^2 M f_\pi^4} \frac{\vec{\sigma}_1 \cdot \vec{q}\, \vec{\sigma}_2 \cdot \vec{q}}{(m_\pi^2 + q^2)^3}}
\end{equation}

can be obtained for self-closings, which again, is linear in the neutron density $\rho_n = k_n^3/3\pi^2$. \\
On the other hand, the two types of pionic vertex corrections give rise to

\begin{equation}\boxed{
	\begin{aligned}
		V_{med}^{(1)} \ \ = \ \ & \frac{g_A^4\, q^2}{64\pi^2 M f_\pi^4} \frac{\vec{\sigma}_1 \cdot \vec{q}\, \vec{\sigma}_2 \cdot \vec{q}}{(m_\pi^2 + q^2)^2} \left[ \frac{k_n^3}{3} - \frac{m_\pi^2}{2} (\Gamma_0 + \Gamma_1) \right. \\[0.5em]
		& - \left. \frac{q^2}{4} (\Gamma_0 + 3\Gamma_1 + 3\Gamma_3 + \Gamma_5) - 3(\Gamma_2 + \Gamma_4) \right],
	\end{aligned}}
\end{equation}

and

\begin{equation}\boxed{
	\begin{aligned}
		V_{med}^{(2)} \ \ = \ \ & \frac{g_A^4}{64\pi^2 M f_\pi^4} \frac{\vec{\sigma}_1 \cdot \vec{q}\, \vec{\sigma}_2 \cdot \vec{q}}{m_\pi^2 + q^2} \left\{ \frac{8k_n^3}{3} + 4m_\pi^2 (\gamma_2 + m_\pi^2 \gamma_0 - 2\Gamma_0) - 4\Gamma_2 \right. \\[0.5em]
		& + q^2 \Big[ m_\pi^2 (\gamma_0 + 2\gamma_1 + \gamma_3) + 2(\gamma_2 + \gamma_4) - \Gamma_0 - 2\Gamma_1 - \Gamma_3 \Big] \\[0.5em]
		& + (8p^2 - q^2) \left[ \frac{q^2}{4} (\gamma_0 + 3\gamma_1 + 3\gamma_3 + \gamma_5) + \gamma_2 + \gamma_4 \right] \\[0.5em]
		& \left. + \left( 8p^2 - \frac{q^2}{2} \right) \Big[ m_\pi^2 (\gamma_0 + \gamma_1) - \Gamma_0 - \Gamma_1 \Big] \right\}.
	\end{aligned}}
\end{equation}

Lastly, only the contribution due to double exchanges remains, which for the sake of clarity and readability, is divided into the results arising from the first ($\,V_{med}^{(3')}\,$) and second line ($\,V_{med}^{(3'')}\,$) of eq. (\ref{eq:r2p2}), respectively.\\
After some tedious calculations, one obtains:

\begin{equation}\boxed{
	\begin{aligned}
		V_{med}^{(3')} \ \ = \ \ & \frac{g_A^4}{128\pi^2 M f_\pi^4} \left\{ \frac{q^2}{4} (6\Gamma_1 - 4\gamma_2 - q^2\gamma_3) - \frac{4k_n^3}{3} + \frac{3}{4} (2m_\pi^2 + q^2) \big( 4\Gamma_0 - q^2\gamma_1 \big) \right. \\[0.5em]
		& -  \frac{3}{4} (2m_\pi^2 + q^2)^2 \big( \gamma_0 + G_0 \big) + \frac{1}{8} (2m_\pi^2 + q^2)^3 K_0 \\[0.5em]
		& + \left[ \frac{q^2}{4} (\gamma_1 + \gamma_3) - \Gamma_0 - \Gamma_1 + \left(m_\pi^2 + \frac{q^2}{2} \right) (\gamma_0 + \gamma_1 + G_0 + 2G_1) \right. \\[0.5em]
		& - \left. \left. \frac{1}{8} (2m_\pi^2 + q^2)^2 \big( K_0 + 2K_1 \big) \right]  i\, (\vec{\sigma}_1 + \vec{\sigma}_2) \cdot (\vec{q} \cross \vec{p})    \vphantom{\frac{q^2}{2}}\right\},
	\end{aligned}}
\end{equation}

and

\begin{equation}\boxed{
	\begin{aligned}
		V_{med}^{(3'')} \ \ = \ \ & \frac{g_A^4}{128\pi^2 M f_\pi^4} \left\{ \left[ \vphantom{\frac{q^2}{4}}  \Gamma_0 + \Gamma_1 + 3\gamma_2 + p^2\gamma_3 +5\gamma_4 + p^2\gamma_5 + G_{0*} + 2G_{1*} \right. \right. \\[0.5em]
		& - \frac{q^2}{4} (\gamma_1 + \gamma_3) - \left(m_\pi^2 + p^2 + \frac{q^2}{2} \right) (\gamma_0 + \gamma_1 + G_0 + 2G_1) \\[0.5em]
		& \left. + \left( m_\pi^2 + \frac{q^2}{2} \right) \left( (2m_\pi^2 + q^2 + 4p^2) \frac{K_0 + 2K_1}{4} - K_{0*} - 2K_{1*} \right) \right] \\[0.5em]
		& \times \, i\, (\vec{\sigma}_1 + \vec{\sigma}_2) \cdot (\vec{q} \cross \vec{p}) \\[0.5em]
		& + \left[ (2m_\pi^2 + q^2 + 4p^2) \frac{K_2}{2} - 2K_{2*} - \gamma_2 - G_2 \right] (q^2\, \vec{\sigma}_1 \cdot \vec{\sigma}_2 - \vec{\sigma}_1 \cdot \vec{q}\, \vec{\sigma}_2 \cdot \vec{q}) \\[0.5em]
		& + \left[ (2m_\pi^2 + q^2 + 4p^2) \left( \frac{K_0}{2} + 2K_1 + 2K_3 \right) - \gamma_0 - 2\gamma_1 - \gamma_3 \right. \\[0.5em]
		& - \left.\left.\vphantom{\frac{K_0}{2}} G_0 - 4G_1 - 4G_3 - 2 (K_{0*} + 4K_{1*} + 4K_{3*}) \right]\, \vec{\sigma}_1 \cdot (\vec{q} \cross \vec{p})\, \vec{\sigma}_2 \cdot (\vec{q} \cross \vec{p}) \right\}.
		\end{aligned}}
	\label{eq:r2p232}
\end{equation}

The present calculation has revealed a typing error $G_{0*} + G_{1*}$ $\rightarrow$ $G_{0*} + 2G_{1*}$ in the fifth line of eq. (45) in ref. \cite{sr} that is corrected in eq. (\ref{eq:r2p232}) above. \\
Let it be noted that once again, the quadratic spin-orbit operator $\vec{\sigma}_1 \cdot (\vec{q} \cross \vec{p})\, \vec{\sigma}_2 \cdot (\vec{q} \cross \vec{p})$ appears directly in the course of the calculation in both of the contributions above. \\
The newly introduced loop-functions $K_\nu,\, \nu = 0, \dots , 3$ are defined by Fermi sphere integrals over the symmetric sum of the product of two different pion propagators, one of which is squared. The $K_\nu$ are discussed in more detail in the appendix, and they also exhibit an additional dependence on the momentum-transfer $q$\,. Lastly, a new notation is employed, whereby an additional factor of $l^2$ in the Fermi sphere integral is denoted by an asterisk $*$ in the subscript of the loop-function. Here, $l$ stands for the 3-momentum modulus of neutrons in the Fermi sea. \\\

This concludes the chapter on effective in-medium nn-potentials, resulting from closing one neutron line in the short-range terms and relativistic corrections to the 3n-force at N$^3$LO in Chiral Effective Field Theory. \\
In the next chapter, a summarizing conclusion is given, followed by an outlook to further research, in order to build upon the results of this and previous works in nuclear many-body calculations.

\chapter{Conclusion and Outlook}\label{chapter:conclusion}

At its starting point, this work deals with three-nucleon forces of Chiral Effective Field Theory, which are crucial for achieving an accurate description of nuclear phenomena. Specifically, a previously developed method to efficiently include 3N-forces in nuclear many-body computations through a density-dependent potential $V_{med}$ is employed. $V_{med}$ is calculated from the 3N-forces by closing one nucleon line and integrating over the filled Fermi sphere. Making use of this approach, the contributions to an in-medium neutron-neutron interaction representing the corresponding subleading chiral 3n-forces, namely short-range terms and relativistic corrections, have been calculated. \\\

The contributions to $V_{med}$ are given as explicit expressions, some of which depending on loop-functions that are detailed in the appendix, either in closed analytical form, or through a one-parameter radial integral. \\
Thus, it is shown that while many contributions of the in-medium nn-interaction are - apart from a constant factor - identical to the terms in isospin-symmetric matter, some differ drastically, and previously vanishing terms from self-closings now yield non-zero contributions. \\
Therefore, it is evident that in some cases, the contributions to $V_{med}$ in pure neutron matter have to be computed explicitly, while others may be easily adapted from previously calculated terms in isospin-symmetric matter. \\
Consequently, the in-medium nn-potential $V_{med}$ detailed in this work is henceforth available for implementation in e.g. many-body calculations of the equation of state of pure neutron matter. \\\

Nevertheless, there are still some unknowns before moving forward. It is unclear at this stage which terms give rise to significant alterations to ordinary two-body nn-interactions, and which are possibly negligible. Therefore, a detailed partial-wave analysis of $V_{med}$ has yet to be carried out, in order to determine the size of its contributions. \\
Lastly, since this work only deals with the short-range terms and relativistic corrections at N$^3$LO, the computations of $V_{med}$ from the intermediate- and long-range contributions at the same order, as well as higher order interaction terms of the chiral 3n-forces, still have to be performed in the future.

\begin{appendices}
\renewcommand{\appendixpagename}{Appendix}
\appendixpage
\renewcommand\thesection{\arabic{section}}

\chapter{Loop-Functions}\label{chapter:loop_functions}

On the following pages, the loop-functions $\Gamma_\nu (p,\,k_n)$,\, $\gamma_\nu (p,\,k_n)$,\, $\nu = 1, \dots, 5$\, and $G_\nu (p,\,q,\,k_n)$,\, $K_\nu (p,\,q,\,k_n)$,\, $\nu = 1, 2, 3$\, used previously to write down the in-medium potentials $V_{med}$\,, are specified. The dependencies of $\Gamma_\nu\,,\, \gamma_\nu$ on $p\,,\, k_n$\,, and the dependencies of $G_\nu\,,\, K_\nu$ on $p\,,\, q\,,\, k_n$ are suppressed for the sake of notational simplicity. Furthermore, the relevant integrals are shown as a function of $\vec{p}$, but yield the same results when computed as a function of $\vec{p}\,'$, since only on-shell scattering with $|\,\vec{p}\,| = p = |\,\vec{p}\,'\,|$ is considered. \\
The loop-functions are obtained by employing the methods detailed in section (\ref{sec:method}) of chapter \ref{chapter:topologies}, most notably by making use of the symmetry under the exchange of three-momentum indices.

\section{$\Gamma_\nu$ Functions}

The $\Gamma_\nu$ functions - with $\nu = 0, 1, \dots, 5$ - are given by Fermi sphere integrals over a single pion propagator with the additional tensorial factors of 1 \ ($\nu = 0$),\, $l_i$ \ ($\nu = 1$), \, $l_i l_j$ \ ($\nu = 2, 3$) \ and \ $l_i l_j l_k$ \ ($\nu = 4, 5$), where \ $i, j, k = 1, 2, 3$. The decompositions of the relevant integrals read:

\begin{equation}
	\begin{aligned}
    	\int\limits_{|\,\vec{l}\,|\, < \,k_n} \frac{d^3l}{2\pi} \, \frac{\{1,\, l_i,\, l_i l_j,\, l_i l_j l_k\}}{m_\pi^2 + (\,\vec{l} + \vec{p}\,)^2} \ \ = \ \ & \Big\{\Gamma_0,\, p_i\,\Gamma_1,\, \delta_{ij}\,\Gamma_2 + p_ip_j\,\Gamma_3, \\
    	& (p_i \delta_{jk} + p_j \delta_{ik} + p_k \delta_{ij})\,\Gamma_4 + p_ip_jp_k\,\Gamma_5 \Big\}\,,
    \end{aligned}
\end{equation}

leading to the following ($p,\, k_n$)-dependent functions

\begin{equation}
	\begin{aligned}
    	\Gamma_0 \ \ = \ \ & k_n - m_\pi \left [\arctan\frac{k_n + p}{m_\pi} + \arctan\frac{k_n - p}{m_\pi}\right] + \frac{m_\pi^2 + k_n^2 - p^2}{4p} \\[0.5em]
    	& \times \ln\frac{m_\pi^2 + (k_n + p)^2}{m_\pi^2 + (k_n - p)^2}\,,
    \end{aligned}
\end{equation}

\begin{equation}
	\begin{aligned}
    	\Gamma_1 \ \ = \ \ & \frac{k_n}{4p^2}(m_\pi^2 + k_n^2 + p^2) - \Gamma_0 - \frac{1}{16p^2}[m_\pi^2 + (k_n + p)^2][m_\pi^2 + (k_n - p)^2] \\[0.5em]
    	& \times \ln\frac{m_\pi^2 + (k_n +p)^2}{m_\pi^2 + (k_n - p)^2}\,,
    \end{aligned}
\end{equation}

\begin{equation}
    \Gamma_2 \ \ = \ \ \frac{k_n^3}{9} + \frac{1}{6}(k_n^2 - m_\pi^2 - p^2)\Gamma_0 + \frac{1}{6}(m_\pi^2 + k_n^2 - p^2)\Gamma_1\,,
\end{equation}

\begin{equation}
    \Gamma_3 \ \ = \ \ \frac{k_n^3}{3p^2} - \frac{m_\pi^2 + k_n^2 + p^2}{2p^2}\Gamma_0 - \frac{m_\pi^2 + k_n^2 + 3p^2}{2p^2}\Gamma_1\,,
\end{equation}

\begin{equation}
	\begin{aligned}
    	\Gamma_4 \ \ = \ \ & \frac{m_\pi^2}{3}\Gamma_0 + \frac{k_n}{64}\left[\frac{5p^2}{3} - 3m_\pi^2 - \frac{31k_n^2}{9} + \frac{1}{3p^2}(3k_n^4 - 14k_n^2m_\pi^2 - 17m_\pi^4) \right. \\[0.5em]
    	& - \left. \frac{(k_n^2 + m_\pi^2)^3}{p^4}\right] + \frac{1}{768p^5}[m_\pi^2 + (k_n + p)^2][m_\pi^2 + (k_n - p)^2] \\[0.5em]  & \times \left[3(k_n^2+m_\pi^2)^2 + 2p^2(k_n^2 + 7m_\pi^2) - 5p^4\right] \ln\frac{m_\pi^2 + (k_n+p)^2}{m_\pi^2 + (k_n-p)^2}\,,
    \end{aligned}	
\end{equation}

\begin{equation}
	\begin{aligned}
    	\Gamma_5 \ \ = \ \ & -\Gamma_0 + \frac{k_n}{64} \left[29 + \frac{5}{p^6}(k_n^2 + m_\pi^2)^3 + \frac{25k_n^2 + 141m_\pi^2}{3p^2} \right. \\[0.5em]
    	& + \left. \frac{1}{3p^4}(17k_n^4 + 86k_n^2m_\pi^2 + 69m_\pi^4) \right] - \frac{1}{256p^7}\big[m_\pi^2 + (k_n + p)^2\big] \\[0.5em]
    	& \times \big[m_\pi^2 + (k_n - p)^2\big] \left[5(k_n^2 + m_\pi^2)^2 + 2p^2(7k_n^2 + 9m_\pi^2) + 29p^4 \right] \\[0.5em]
    	& \times \ln\frac{m_\pi^2 + (k_n +p)^2}{m_\pi^2 + (k_n - p)^2}\,.
   	\end{aligned}
\end{equation}

\section{$\gamma_\nu$ Functions}

The $\gamma_\nu$ functions - with $\nu = 0, 1, \dots, 5$ - are given by Fermi sphere integrals over a squared pion propagator with the additional tensorial factors of 1 \ ($\nu = 0$),\, $l_i$ \ ($\nu = 1$),\, $l_i l_j$ \ ($\nu = 2, 3$) \ and \ $l_i l_j l_k$ \ ($\nu = 4, 5$), where \ $i, j, k = 1, 2, 3$. The decompositions of the relevant integrals read:

\begin{equation}
	\begin{aligned}
		\int\limits_{|\,\vec{l}\,| \,<\, k_n} \frac{d^3l}{2\pi} \, \frac{\{1,\, l_i,\, l_i l_j,\, l_i l_j l_k\}}{\big[m_\pi^2 + (\,\vec{l} + \vec{p}\,)^2\big]^2}\ \ = \ \ & \Big\{\gamma_0,\, p_i\,\gamma_1,\, \delta_{ij}\,\gamma_2 + p_ip_j\,\gamma_3, \\
		& (p_i \delta_{jk} + p_j \delta_{ik} + p_k \delta_{ij})\,\gamma_4 + p_ip_jp_k\,\gamma_5 \Big\}.
	\end{aligned}
\end{equation}

Obviously, the relation $\gamma_\nu = - \partial \Gamma_\nu / \partial m_\pi^2$ is fulfilled by definition. With that knowledge, the analytical form of these ($p,\, k_n$)-dependent loop-functions is obtained as

\begin{equation}
	\begin{aligned}
		\gamma_0 \ \ = \ \ & \frac{1}{2m_\pi} \left[\arctan\frac{k_n + p}{m_\pi} + \arctan\frac{k_n - p}{m_\pi} \right] - \frac{1}{4p} \ln\frac{m_\pi^2 + (k_n + p)^2}{m_\pi^2 + (k_n - p)^2}\,,
	\end{aligned}
\end{equation}

\begin{equation}
	\begin{aligned}
		\gamma_1 \ \ = \ \ & -\gamma_0 - \frac{k_n}{2p^2} + \frac{p^2 + k_n^2 + m_\pi^2}{8p^3} \ln\frac{m_\pi^2 + (k_n + p)^2}{m_\pi^2 + (k_n - p)^2}\,,
	\end{aligned}
\end{equation}

\begin{equation}
	\begin{aligned}
		\gamma_2 \ \ = \ \ & \frac{k_n}{8p^2}(3p^2-k_n^2-m_\pi^2) - m_\pi^2\gamma_0 + \frac{1}{32p^3}\big[(p^2 + k_n^2 + m_\pi^2)^2 \\[0.5em]
		& - 4p^2 (p^2 + m_\pi^2)\big] \ln\frac{m_\pi^2 + (k_n + p)^2}{m_\pi^2 + (k_n - p)^2}\,,
	\end{aligned}
\end{equation}

\begin{equation}
	\begin{aligned}
		\gamma_3 \ \ = \ \ & \gamma_0 + \frac{k_n}{8p^4}(7p^2 + 3k_n^2 + 3m_\pi^2) -  \frac{1}{32p^5}\big[3(k_n^2 + m_\pi^2)^2 \\[0.5em]
		& +2p^2(3k_n^2 + 5m_\pi^2)^2 +7p^4\big]\ln\frac{m_\pi^2 + (k_n + p)^2}{m_\pi^2 + (k_n - p)^2}\,,
	\end{aligned}
\end{equation}

\begin{equation}
	\begin{aligned}
		\gamma_4 \ \ = \ \ & m_\pi^2\gamma_0 + \frac{k_n}{16p^4} \left[(k_n^2 + m_\pi^2)^2 + \frac{4p^2}{3}(k_n^2 + 3m_\pi^2) - 5p^4 \right] \\[0.5em]
		& + \frac{p^2- k_n^2 - m_\pi^2}{64p^5}\big[5p^4 + 2p^2(k_n^2 + 3m_\pi^2)  + (k_n^2 + m_\pi^2)^2\big] \\[0.5em]
		& \times \ln\frac{m_\pi^2 + (k_n + p)^2}{m_\pi^2 + (k_n - p)^2}\,,
	\end{aligned}
\end{equation}

\begin{equation}
	\begin{aligned}
		\gamma_5 \ \ = \ \ & -\gamma_0 - \frac{k_n}{p^2} \left[ \frac{5}{16p^4}(k_n^2 + m_\pi^2)^2 + \frac{19}{16} + \frac{2k_n^2 + 3m_\pi^2}{3p^2} \right] \\[0.5em]
		& + \frac{1}{64p^7}\big[5p^4(3k_n^2 + 7m_\pi^2) + 19p^6 + 3p^2(k_n^2 + m_\pi^2)(3k_n^2 + 7m_\pi^2) \\[0.5em]
		& + 5(k_n^2 + m_\pi^2)^3\big] \ln\frac{m_\pi^2 + (k_n + p)^2}{m_\pi^2 + (k_n - p)^2}\,.
	\end{aligned}
\end{equation}

\section{$G_\nu$ Functions}

The $G_\nu$ functions - with $\nu = 0, 1, 2, 3$ - are given by Fermi sphere integrals over two different pion propagators with the additional tensorial factors of 1 \ ($\nu = 0$),\, $l_i$ \ ($\nu = 1$) and \ $l_i l_j$ \ ($\nu = 2, 3$), where \ $i, j = 1, 2, 3$. The decompositions of the relevant integrals read:

\begin{equation}
	\begin{aligned}
		\int\limits_{|\,\vec{l}\,| \,<\, k_n} \frac{d^3l}{2\pi} \, \frac{\{1,\, l_i,\, l_i l_j\}}{\big[m_\pi^2 + (\,\vec{l} + \vec{p}\,)^2\big]\big[m_\pi^2 + (\,\vec{l} + \vec{p}\,'\,)^2\big]}\ \ = \ \ & \Big\{G_0,\, (p_i + p_i')\,G_1, \\
		& \, \delta_{ij}\,G_2 + (p_i + p_i')(p_j + p_j')\,G_3 \\[0.5em]
		& + (p_i - p_i')(p_j - p_j')\,G_4 \Big\},
	\end{aligned}
\end{equation}

where the symmetry under $\vec{p} \leftrightarrow \vec{p}\,'$ has been exploited. Although it is required for the construction of $G_2$ and $G_3$\,, the function $G_4$ itself does not appear in the contributions to the in-medium nn-potential $V_{med}$ and does not need to be specified further. \\
However, contrary to the previous loop functions, the integrals cannot be solved analytically. Instead, they are obtained by first computing the radial one-parameter integrals

\begin{equation}
	G_{0,\, 0*,\, **} \ \ = \ \ \frac{2}{q}\int\limits_0^{k_n} dl \, \frac{\{l,\, l^3,\, l^5\}}{\sqrt{B+q^2l^2}}\, \ln\frac{q\,l + \sqrt{B + q^2l^2}}{\sqrt{B}}
\end{equation}

numerically, where $B = \big[m_\pi^2 + (\,\vec{l} + \vec{p}\,)^2\big]\big[m_\pi^2 + (\,\vec{l} + \vec{p}\,'\,)^2\big]$. The asterisks in the subscript indicate additional factors of $l^2$ in the corresponding Fermi sphere integral. Subsequently, it is necessary to solve a system of linear equations with the result

\begin{align}
	&G_1 \hspace{-1.6cm}&&= \ \ \frac{1}{4p^2-q^2} \big[\Gamma_0 - (m_\pi^2 + p^2)G_0 - G_{0*}\big], \\[0.5em]
	&G_{1*} \hspace{-1.6cm}&&= \ \ \frac{1}{4p^2 - q^2} \big[3\Gamma_2 + p^2\Gamma_3 - (m_\pi^2 + p^2)G_{0*} - G_{**}\big], \\[0.5em]
	&G_2 \hspace{-1.6cm}&&= \ \ (m_\pi^2 + p^2)G_1 + G_{0*} + G_{1*}\,, \\[0.5em]
	&G_3 \hspace{-1.6cm}&&= \ \ \frac{1}{4p^2 - q^2} \left[ \frac{\Gamma_1}{2} - 2(m_\pi^2 + p^2)G_1 - G_{0*} - 2G_{1*} \right].
\end{align}

\section{$K_\nu$ Functions}

The $K_\nu$ functions - with $\nu = 0, 1, 2, 3$ - are given by Fermi sphere integrals over the symmetric sum of a squared pion propagator multiplied by a different pion propagator, with the additional tensorial factors of 1 \ ($\nu = 0$),\, $l_i$ \ ($\nu = 1$) and \ $l_i l_j$ \ ($\nu = 2, 3$), where \ $i, j = 1, 2, 3$. The decompositions of the relevant integrals read:

\begin{equation}
	\begin{aligned}
		&\int\limits_{|\,\vec{l}\,| \,<\, k_n} \frac{d^3l}{2\pi} \,  \left( \frac{1}{\big[m_\pi^2 + (\,\vec{l} + \vec{p}\,)^2\big]^2\big[m_\pi^2 + (\,\vec{l} + \vec{p}\,'\,)^2\big]} + \frac{1}{\big[m_\pi^2 + (\,\vec{l} + \vec{p}\,)^2\big]\big[m_\pi^2 + (\,\vec{l} + \vec{p}\,'\,)^2\big]^2} \right) \\[0.5em]
		& \hphantom{\int\limits_{|\,\vec{l}\,| \,<\, k_n} \frac{d^3l}{2\pi} \,} \times \{1,\, l_i,\, l_i l_j\} \\[0.5em] & = \ \ \Big\{K_0,\, (p_i + p_i')\,K_1,\, \delta_{ij}\,K_2 + (p_i + p_i')(p_j + p_j')\,K_3  + (p_i - p_i')(p_j - p_j')\,K_4 \Big\},
	\end{aligned}
\end{equation}

where the symmetry under $\vec{p} \leftrightarrow \vec{p}\,'$ is used once again. As before, $K_4$ is relevant for the construction of $K_2$ and $K_3$\,, but does not itself appear in the nn-potentials and is thus of no further interest here. \\
Moreover, it can be easily verified that the $K_\nu$ satisfy $K_\nu = - \partial G_\nu / \partial m_\pi^2$\,. Just as for the $G_\nu$ functions, the $K_\nu$ are not analytically calculable and have to be obtained by first computing the radial one-parameter integrals

\begin{equation}
	\begin{aligned}
		K_{0,\, 0*,\, **,\, ***} \ \ = \ \ 2 \int\limits_0^{k_n} dl \, & \frac{m_\pi^2 + l^2 + p^2}{B+q^2l^2} \left[ \frac{l}{B} + \frac{1}{q \sqrt{B + q^2l^2}}\, \ln\frac{q\,l + \sqrt{B + q^2l^2}}{\sqrt{B}} \right] \\[0.5em]
		& \times \{l,\, l^3,\, l^5, l^7\}
		\end{aligned}
\end{equation}

numerically, and then solving a system of linear equations with the result

\begin{align}
	&K_1 \hspace{-3mm}&&= \ \ \frac{1}{4p^2 - q^2} \big[\gamma_0 + G_0 - (m_\pi^2 + p^2)K_0 - K_{0*}\big], \\[0.5em]
	&K_{1*} \hspace{-3mm}&&= \ \ \frac{1}{4p^2 - q^2} \big[3\gamma_2 + p^2\gamma_3 + G_{0*} - (m_\pi^2 + p ^2)K_{0*} - K_{**}\big], \\[0.5em]
	&K_2 \hspace{-3mm}&&= \ \ (m_\pi^2 + p ^2)K_1 - G_1 + K_{0*} + K_{1*}\,, \\[0.5em]
	&K_3 \hspace{-3mm}&&= \ \ \frac{1}{4p^4 - q^2} \left[ \frac{\gamma_1}{2} - 2(m_\pi^2 + p^2)K_1 + 2G_1 - K_{0*} - 2K_{1*} \right], \\[0.5em]
	&K_{1**} \hspace{-3mm}&&= \ \ \frac{1}{4p^2 - q^2} \big[G_{**} - (m_\pi^2 + p^2)K_{**} - K_{***} + \gamma_{**}\big], 
	\label{eq:gammastst} \\[0.5em]
	&K_{2*} \hspace{-3mm}&&= \ \ (m_\pi^2 + p^2)K_{1*} - G_{1*} + K_{**} + K_{1**}\,, \\[0.5em]
	&K_{3*} \hspace{-3mm}&&= \ \ \frac{1}{4p^2 - q^2} \left[ \frac{1}{2} (5\gamma_4 + p^2\gamma_5) - 2(m_\pi^2 + p^2)K_{1*} + 2G_{1*} - K_{**} - 2K_{1**}  \right].
\end{align}

For the sake of simplicity, a new loop function $\gamma_{**}$ - defined as the Fermi sphere integral over a squared pion propagator multiplied by $l^4$ - is introduced in eq. (\ref{eq:gammastst}). \\
Its analytical form reads:

\begin{equation}
	\begin{aligned}
		\gamma_{**} \ \ &= \ \ \int\limits_{|\,\vec{l}\,| \,<\, k_n} \frac{d^3l}{2\pi} \, \frac{l^4}{\big[m_\pi^2 + (\,\vec{l} + \vec{p}\,)^2\big]^2}\\[0.5em]
		&= \ \ (p^4 + 5m_\pi^4 - 10p^2m_\pi^2)\gamma_0 + 4k_n \left( p^2 + \frac{k_n^2}{6} - m_\pi^2 \right) + \frac{m_\pi^4 - p^4}{p} \\[0.5em]
		&\hphantom{=} \ \ \times \ln\frac{m_\pi^2 + (k_n + p)^2}{m_\pi^2 + (k_n - p)^2}\,.
	\end{aligned}
\end{equation}

\end{appendices}

\microtypesetup{protrusion=false}
\microtypesetup{protrusion=true}
\nocite{*}
\printbibliography[
title={References},
heading=bibintoc]

\end{document}